\documentclass[proof]{pasj00}
\usepackage{lscape}
\usepackage{multirow}
\usepackage{bigdelim}

\SetRunningHead{N.~Iso \etal}{Origin of the  Broad  Iron Line Feature and the Soft X-ray Variation in Seyfert  Galaxies }

\Received{\today}
\Accepted{\today}

\begin{document}
\title{Origin of the  Broad Iron Line Feature and the Soft X-ray Variation in  Seyfert  Galaxies} 
\author{%
Naoki~\textsc{Iso}\altaffilmark{1,2,3},
Ken~\textsc{Ebisawa}\altaffilmark{1,2}, Hiroaki Sameshima\altaffilmark{1}, Misaki Mizumoto\altaffilmark{1,2},\\
Takehiro Miyakawa\altaffilmark{4}, Hajime Inoue\altaffilmark{1,5} and Hiroki Yamasaki\altaffilmark{1,2}
}
\altaffiltext{1}{%
Institute of Space and Astronautical Science (ISAS), \\Japan Aerospace Exploration Agency (JAXA)\\  
3-1-1 Yoshinodai, Chuo-ku, Sagamihara, Kanagawa 252-5210}
\altaffiltext{2}{%
Department of Astronomy, Graduate School of Science, The University of Tokyo,
7-3-1 Hongo, \\ Bunkyo-ku, Tokyo 113-0033}
\altaffiltext{3}{Tokyo Seitoku University High School, 6-7-14 Oji, Kita-ku, Tokyo, 114-0002}
\altaffiltext{4}{Tsukuba Space Center (TKSC), Japan Aerospace Exploration Agency (JAXA),
2-1-1 Sengen, \\ Tsukuba-shi, Ibaraki 305-8505}
\altaffiltext{5}{Meisei University, 2-1-1 Hodokubo, Hino,  Tokyo 191-8506}
\email{ebisawa@isas.jaxa.jp}
%
\maketitle

\begin{abstract}
Many Seyfert  galaxies are known to exhibit significant X-ray spectral variations and seemingly broad iron K-emission line features.
In this paper, we show that the ``variable partial covering model'', which has been successfully proposed
for MCG-6-30-15 (\cite{Miyakawa2012}) and  1H0707--495 (Mizumoto, Ebisawa \& Sameshima 2014), can also 
explain the spectral variations in 2--10 keV as well as the broad iron line features
in  20 other Seyfert  galaxies observed with Suzaku.
In this model, the absorbed spectral component through the optically-thick absorbing clouds has  a significant
 iron K-edge, which primarily  accounts for the observed seemingly broad iron line feature.  Fluctuation  of the absorbing
clouds in the line of sight of  the extended X-ray source results in variation  of the
partial covering fraction, which causes   an anti-correlation between the 
 direct (not covered)  spectral component  and the absorbed  (covered) spectral  component below $\sim$10 keV.
Observed spectral variation in 2--10 keV in a timescale of 
less than  $\sim$day is primarily explained  by such variations of the partial covering fraction,
while the intrinsic soft X-ray  luminosity is hardly variable. 

\end{abstract}

\section{Introduction}\label{s1}
Significant and aperiodic 
X-ray  variation  is one of the main characteristics of the 
Active Galactic Nuclei (AGN).
However, origin of the AGN X-ray variation  has not been fully understood yet, 
in spite of  extensive observational and theoretical studies
(e.g., \cite{Mus,Ulr}).

%
%

In particular, many Seyfert  galaxies are known to exhibit an  intriguing spectral
variation  in the iron K-energy band ($6 \sim 7$ keV).
MCG-6-30-15 is a representative example; 
its iron emission line profile seems to be broadened
and skewed  (e.g., \cite{Tanaka1995}), and fractional variation of the energy spectrum 
significantly drops at the iron line energy band \citep{Fabian2002,Matsumoto2003}.  A possible scenario to
explain these phenomena  is the ``light-bending model''.  In this model, 
the fluorescent iron line is emitted at the innermost part of the accretion disk, 
so that the  line profile is broadened and skewed, and 
the disk-reflected photons are much less variable than the  direct photons due to relativistic 
reverberation \citep{FandV,Miniutti2004}.
Alternatively, the seemingly broad iron emission line feature may be interpreted due to 
iron K-edge feature caused by 
partial covering of the central X-ray source
by intervening absorbers in the line of sight (e.g., \cite{Matsu,Inoue2003,Miller2008}).
In this model, the 
apparent invariability of the iron energy band is explained as due to such 
relatively higher variation of the continuum that is   caused by change of the  amounts of the absorption \citep{Inoue2003}.

 \citet{Miyakawa2012}  (MEI2012 hereafter)
 proposed a ``variable partial covering (VPC) model'',
which is a sophistication of the latter idea.
In this model, below $\sim$10 keV, the original X-ray luminosity of the AGN is not significantly
variable,   and the
apparent X-ray variation is primarily caused by 
variation of the geometrical covering fraction of the extended X-ray source 
by the intervening clouds having internal ionization structures.
MEI2012  applied the VPC model to Suzaku observations of 
MCG-6-30-15, and successfully explained not
only the small fractional variation in the iron energy band, but also the spectral variations 
in 1 -- 10 keV (see also \cite{IME}), besides an independent hard-tail variations above $\sim$10 keV. 

Mizumoto, Ebisawa \& Sameshima (2014) (MES2014 hereafter) has shown that  the VPC model can also explain the soft X-ray spectral variations of the narrow Seyfert 1 galaxy 1H0707--405
 observed with XMM and Suzaku below $\sim$10 keV. Therefore, 
it is interesting to examine if  the VPC model is  valid for 
other Seyfert  galaxies to explain the spectral variation below $\sim$10 keV and the broad iron emission line feature.
In the present paper, we explore the Suzaku archive to select
Seyfert  galaxies that show similar X-ray spectral characteristics to 
MCG-6-30-15,  and  apply the VPC model to see if  their X-ray 
spectral shapes and variations in 2--10 keV are explained by this model.

\section{Observations and Data Reduction}\label{s2}
\begin{table*}[t]
\begin{center}
\begin{center}
{
\renewcommand\arraystretch{1.}
	\caption{Log of the observations.}
	\label{t1}
\begin{tabular}{ccccccccc}
\hline
Observation & \phantom{OO}Target		&	$\alpha$		&	$\delta$		&	${\it z}$	&	Sequence		&
Start					&	$t_{\rm exp}$	&						XIS			\\
\#&					&	\multicolumn{2}{c}{(J2000.0)}					&&	\#			&
Date					&	(ks)			&									\\
\hline
\multicolumn{9}{c}{Type I Seyfert} \\
\hline
1&Markarian\,335			&	00:06:19.582	&	+20:12:10.58	&	0.0254	&	701031010	&
2006-06-21			&	151						&	 	0 2 3	\\
\hline
2&TonS\,180				&	00:57:19.940	& 	$-$22:22:59.10	&	0.0620	&	701021010	&
2006-12-09			&	120						&		0 \phantom{2} 3\\
\hline
3&1H\,0323+342 			&	03:24:41.161	&	+34:10:45.86	&	0.0629	&	704034010	&
2009-07-26			&	\phantom{1}84	 			&		0 \phantom{2} 3\\
\hline
4&NGC\,1365			&	03:33:36.310	&	$-$36:08:27.80&	0.0056	&	705031010	&
2010-06-27			&	151							&	0 \phantom{2} 3  \\
&					& 				&				&			&	705031020	&
2010-07-15			&	302							&	0 \phantom{2} 3\\
\hline
5&3C\,111			&	04:18:21.277	&	+38:01:35.80	&	0.0485	&	705040010	&
2010-09-02			&	\phantom{1}80				&		0 \phantom{2} 3  \\
&					&				&				&			&	705040020	&
2010-09-09			&	\phantom{1}79				&		0 \phantom{2} 3  \\
&					& 				&				&			&	705040030	&
2010-09-14			&	\phantom{1}80				&		0 \phantom{2} 3\\
\hline
6&1H\,0419--577 (1)		&	04:26:00.715	&	$-$57:12:01.69	&	0.1040	&	702041010	&
2007-07-25			&	205								&	0 \phantom{2} 3\\
\hline
7&1H\,0419--577 (2)		&	--			&	--			&	--		&	704064010	&
2010-01-16			&	123								&	0 \phantom{2} 3\\
\hline
8&Arakelian\,120			&	05:16:11.395	&	$-$00:08:59.65	&	0.0323	&	702014010	&
2007-04-01			&	100								&	0 \phantom{2} 3\\
\hline
9&NGC\,3227		&	10:23:30.608	&	+19:51:53.82	&	0.0037	&	703022010	&
2008-10-28			&	\phantom{1}58				&		0 \phantom{2} 3\\
&					& 				&				&			&	703022020	&
2008-11-04			&	\phantom{1}53				&		0 \phantom{2} 3\\
&					&				&				&			&	703022030	&
2008-11-12			&	\phantom{1}56				&		0 \phantom{2} 3\\
&					&				&				&			&	703022040	&
2008-11-20			&	\phantom{1}64				&		0 \phantom{2} 3\\
&					&				&				&			&	703022050	&	
2008-11-27			&	\phantom{1}79				&		0 \phantom{2} 3\\
&					&				&				&			&	703022060	&
2008-12-02			&	\phantom{1}51				&		0 \phantom{2} 3\\
\hline
10&NGC\,3516			&	11:06:47.494	&	+72:34:06.70	&	0.0088	&	704062010	&
2009-10-28			&	251								&	0 \phantom{2} 3\\
\hline
11&NGC\,3783 (1)			&	11:39:01.721	&	$-$37:44:18.60	&	0.0097	&	701033010	&
2006-06-24			&	\phantom{1}75				&	 0 2 3\\
\hline
12&NGC\,3783 (2)			&	--			&	--			&	--		&	704063010	&
2009-07-10			&	210								&	0 \phantom{2} 3\\
\hline
13&NGC\,4051		&	12:03:09.686	&	+44:31:52.54	&	0.0022	&	703023010	&
2008-11-06			&	274								&	0 \phantom{2} 3\\
&					&				&				&			&	703023020	&
2008-11-23			&	\phantom{1}78				&		0 \phantom{2} 3\\
\hline
14&NGC\,4151			&	12:10:32.659	&	+39:24:20.74	&	0.0033	&	701034010	&
2006-12-18			&	124								&	0 \phantom{2} 3\\
\hline
15&Markarian\,766			&	12:18:26.484	&	+29:48:46.15	&	0.0123	&	701035010	&
2006-11-16			&	\phantom{1}98				&		0 \phantom{2} 3\\
\hline
16&Markarian\,205			&	12:21:43.967	&	+75:18:37.99	&	0.0708	&	705062010	&
2010-05-22			&	100								&	0 \phantom{2} 3\\
\hline
17&NGC\,4593			&	12:39:39.492	&	$-$05:20:39.16	&	0.0090	&	702040010	&
2007-12-15			&	118								&	0 \phantom{2} 3\\
\hline
 \multicolumn{9}{@{}l@{}}{\hbox to 0pt{\parbox{180mm}{\footnotesize 
  }\hss}}
\end{tabular}
}
\end{center}
\end{center}
\end{table*}
\addtocounter{table}{-1}
\begin{table*}[t]
\begin{center}
\begin{center}
{
\renewcommand\arraystretch{1.}
	\caption{Log of the observations (continued)}
	\label{t1}
\begin{tabular}{ccccccccc}
\hline
Observation &\phantom{OO}Target	&	$\alpha$		&	$\delta$		&	${\it z}$	&	Sequence		&
Start					&	$t_{\rm exp}$	&						XIS			\\
\#&					&	\multicolumn{2}{c}{(J2000.0)}		&			&	\#			&
Date					&	(ks)			&									\\
\hline
\multicolumn{9}{c}{Type I Seyfert} \\
\hline
18&IC\,4329A		&	13:49:19.277	&	$-$30:18:33.83	&	0.0160	&	702113010	&
2007-08-01			&	\phantom{1}25				&		0 \phantom{2} 3\\
&					&				&				&			&	702113020	&
2007-08-06			&	\phantom{1}30				&		0 \phantom{2} 3\\
&					&				&				&			&	702113030	&
2007-08-11			&	\phantom{1}26 			&		0 \phantom{2 }3\\
&					&				&				&			&	702113040	&
2007-08-16			&	\phantom{1}24					&	0 \phantom{2} 3\\
&					&				&				&			&	702113050	&
2007-08-20			&	\phantom{1}24						&	0 \phantom{2} 3\\
\hline
19& NGC\,5548		&	14:17:59.513	&	+25:08:12.45	&	0.0165	&	702042010	&
2007-06-18 			&	\phantom{1}31						&	0 \phantom{2} 3\\
&					&				&				&			&	702042020	&
2007-06-24			&	\phantom{1}35						&	0 \phantom{2} 3\\
&					&				&				&			&	702042040	&
2007-07-08			&	\phantom{1}30						&	0 \phantom{2} 3\\
&					&				&				&			&	702042050	&
2007-07-15			&	\phantom{1}30						&	0 \phantom{2} 3\\
&					&				&				&			&	702042060	&
2007-07-22			&	\phantom{1}28						&	0 \phantom{2} 3\\
&					&				&				&			&	702042070	&
2007-07-29			&	\phantom{1}31						&	0 \phantom{2} 3\\
&					&				&				&			&	702042080	&
2007-08-05			&	\phantom{1}38						&	0 \phantom{2} 3\\
\hline
20&4C\,74.26				&	20:42:37.285	&	+75:08:02.36	&	0.1034	&	702057010	&
2007-10-28			&	\phantom{1}91						&	0 \phantom{2} 3  \\
\hline
21&Arakelian\,564			&	22:42:39.309	&	+29:43:31.55	&	0.0249	&	702117010	&
2007-06-26			&	\phantom{1}99						&	0 \phantom{2} 3\\
\hline
22&SWIFT\,J2127.4+5654	&	21:27:45.400	&	+56:56:35.00	&	0.0147	&	702122010	&
2007-12-09			&	\phantom{1}91						&	\phantom{0 2} 3\\
\hline
23&NGC\,7469			&	23:03:15.674	&	+08:52:25.28	&	0.0159	&	703028010	&
2008-06-24			&	112								&	0 \phantom{2} 3\\
\hline
\multicolumn{9}{c}{Type II Seyfert} \\
\hline
24&NGC\,2992		&	09:45:42.045	&	$-$14:19:34.90	&	0.0077	&	700005010	&
2005-11-06			&	\phantom{1}37					&	0 2 3\\
&&				&				&			&	700005020	&
2005-11-19			&	\phantom{1}37						&	0 2 3\\
& &  &  & & 700005030 & 2005-12-13 & \phantom{1}46   & 0 2 3 \\
\hline
25&MCG\,-5-23-16		&	09:47:40.170	&	$-$30:56:55.91	&	0.0082	&	700002010	&
2005-12-07			&	\phantom{1}95						&	0 2 3\\
\hline
26&Centaurus\,A		&	13:25:27.615	&	$-$43:01:08.81	&	0.0018	&	704018010	&
2009-07-20			&	\phantom{1}62						&	0 \phantom{2} 3\\
&					&				&				&			&	704018020	&
2009-08-05			&	\phantom{1}51						&	0 \phantom{2} 3\\
&					&				&				&			&	704018030	& 
2009-08-14			&	\phantom{1}55						&	0 \phantom{2} 3\\
\hline
27&IRAS\,18325--5926		&	18:36:58.257	&	$-$59:24:08.44	&	0.0194	&	702118010	&
2007-10-26			&	\phantom{1}78						&	0 \phantom{2} 3\\
\hline
  \multicolumn{9}{@{}l@{}}{\hbox to 0pt{\parbox{180mm}{\footnotesize 
  \noindent 
  }\hss}}
\end{tabular}
}
\end{center}
\end{center}
\end{table*}
\subsection{Instruments}
The data used in this paper 
 were taken by Suzaku \citep{Mitsuda2007},  which has two operating instruments, the 
X-ray Imaging Spectrometer (XIS; \cite{Koyama2007}) and the Hard X-ray Detector (HXD; \cite{Takahashi2007}; \cite{Kokubun2007}).
The XIS is composed of four CCD cameras, XIS0 to XIS3, 
each of  which is  located on the focal plane of the identical X-ray telescope module (\cite{XRT}).
The XIS is sensitive in 0.2 -- 12.0~keV, and 
the field of view (FOV) is 17$\farcm$8 $\times$ 17$\farcm$8. 
A half power diameter for the  point-spread function is 
 $\sim$2$\arcmin$.  The XIS0, 2, and 3 have front-illuminated (FI) chips, and the 
XIS1 has a back-illuminated (BI) 
one. Since the non-X-ray background (NXB) level of the FI CCDs 
is significantly lower than that of the BI CCD in the iron K-band,
we use only the FI CCDs in the present study. 
The entire XIS2 and a part of the XIS0 are dysfunctional since 2006 November and 2009 June, respectively,
presumably due to  micrometeorite hits. The available XIS FI cameras during each observation are  shown
in the last column of the observation log (Table \ref{t1}).
 The HXD consists of two types of the detectors, PIN and GSO, 
achieving the combined  sensitivity in 10 -- 600~keV.  We use only PIN in this study (10 -- 60 keV), since
our targets are not bright enough to make the GSO spectral study feasible.
The PIN has  a FOV of $\sim$34$\arcmin$ square in FWHM. 

\subsection{Data Selection}\label{ss2-1}
We chose data  from the Suzaku public archive. 
Our main purpose is to study X-ray intensity and spectral
 variations  of the Seyfert galaxies that  have similar spectral 
characteristics to MCG-6-30-15, in particular 
in the iron K-energy band.  Therefore, 
we selected only targets which are   classified as Seyfert galaxies, and 
known to show  the seemingly broad iron K-line feature or a hint of that.

The unit of Suzaku observations is an observation ``sequence'', which is a single continuous pointing period,
typically for a $\sim$day.
For a given target, we combine the  sequences carried out within a $\sim$month  to define 
a new  ``observation'' (Table \ref{t1}). 
To study spectral variations efficiently, we chose only the  observations which satisfy the following conditions;
(1) long enough total exposure  time  ($\gtrsim60$ ksec) for an observation,
(2) bright enough that the total accumulate counts in an observation  are more than 50,000 counts
in 0.2 -- 10 keV,
and  (3) the sources are  variable more than 10\% in 4 -- 10 keV in an observation.
Thus, we selected 27 observations for 25 targets (using 50 sequences in total).
The observation log is shown in Table~\ref{t1}.

\subsection{Data Reduction}\label{ss2-2}

We reprocessed all the data under the standard pipeline version 2.5 and used HEASoft\footnote{See http://heasarc.nasa.gov/docs/software/lheasoft/ for detail.} version 6.11 for  data reduction.
For the XIS, we excluded events obtained during passages through the South Atlantic Anomaly (SAA) and Earth elevation angle $<20^{\circ}$ for the day-time Earth and $<5^{\circ}$ for the night Earth. The source
 events were extracted from a 3$\arcmin$ radius circle centered on the source. The background events were extracted from an annulus of 4$\arcmin$--6$\arcmin$ in radii when the source is located at the XIS nominal position, 
or from a 3$\arcmin$ radius circle offset from the source avoiding the calibration sources at chip corners when the source is located at the 
 HXD nominal position. 
Figure~\ref{f1} shows the 0.2 -- 12.0~keV light curves by XIS for  all the 27 observations.
As for the spectral analysis, we used redistribution matrix functions (RMFs) and ancillary response files (ARFs), created by the \texttt{xisrmfgen} and \texttt{xissimarfgen} \citep{Ishisaki2007} tools. The three or two XIS FI spectra and  responses were combined.

For the HXD/PIN,  we exclude events obtained during passages through the SAA and elevation angle from dark Earth rim $<5^{\circ}$. The PIN background is composed of  the NXB and Cosmic X-ray background (CXB). We simulated background data for the spectral analysis. The NXB spectrum was provided by the instrument team \citep{Fukazawa2009}, while the CXB spectrum was simulated by convolving the HEAO-1 model \citep{Boldt1987} with the detector response.

\section{Data Analysis and Results}\label{s3}
\begin{figure*}
\begin{center}
\includegraphics[trim=3cm 4cm 0.5cm 4cm,clip,width=18cm]{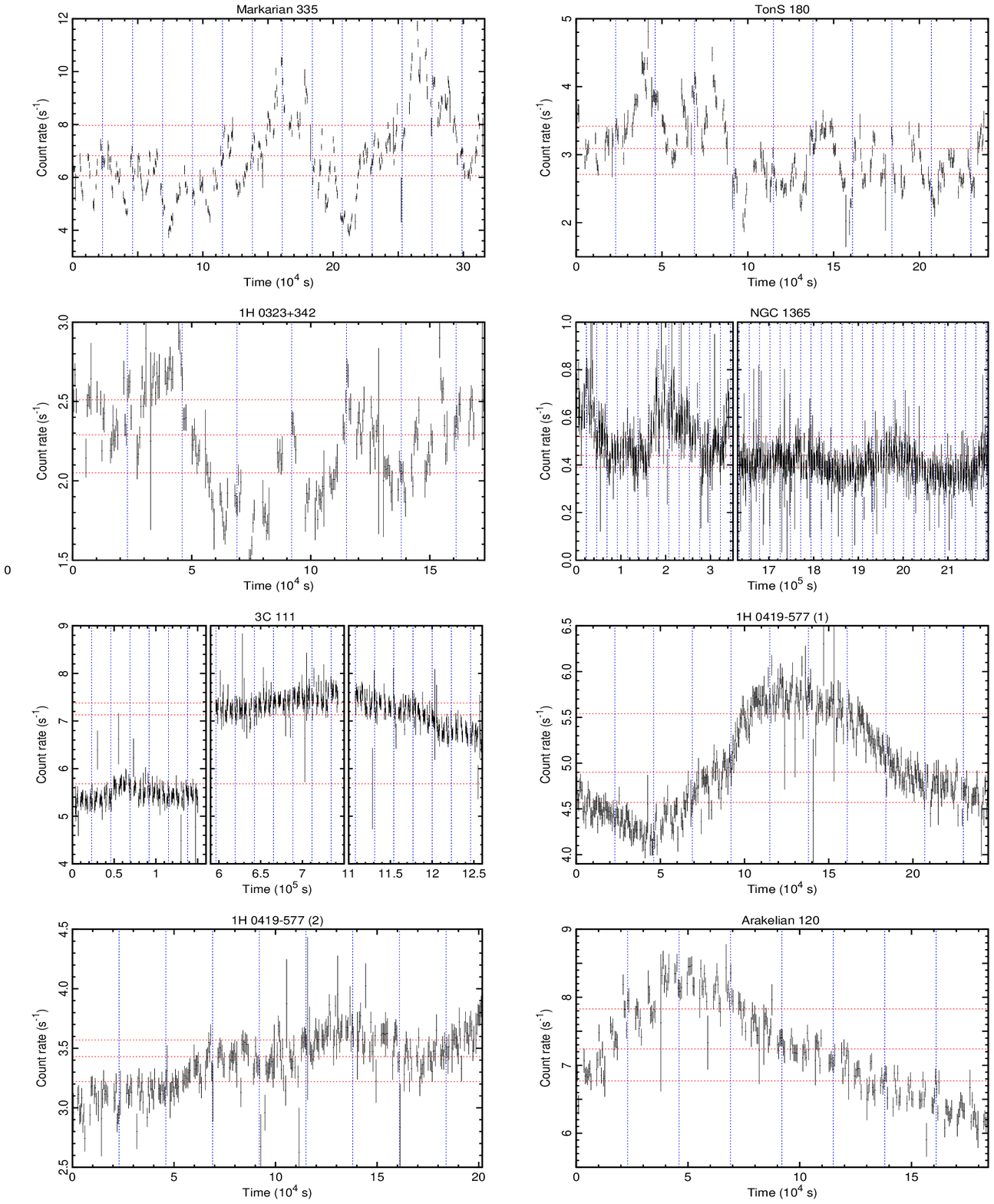}
\end{center}
\caption{XIS light curves of the 27 observations
in 0.2 -- 12.0~keV. The count rate is binned with 512~s. 
Horizontal red-dotted lines show the count-rate intervals with which the intensity-sliced spectra were made
(section \ref{ss3-2}). 
Vertical blue-dotted lines show the time intervals  with which the time-sliced spectra were made
(section \ref{ss3-3}).}
	\label{f1}
\end{figure*}
\addtocounter{figure}{-1}
\begin{figure*}
\begin{center}
\includegraphics[trim=3cm 4cm 0.5cm 4cm,clip,width=18cm]{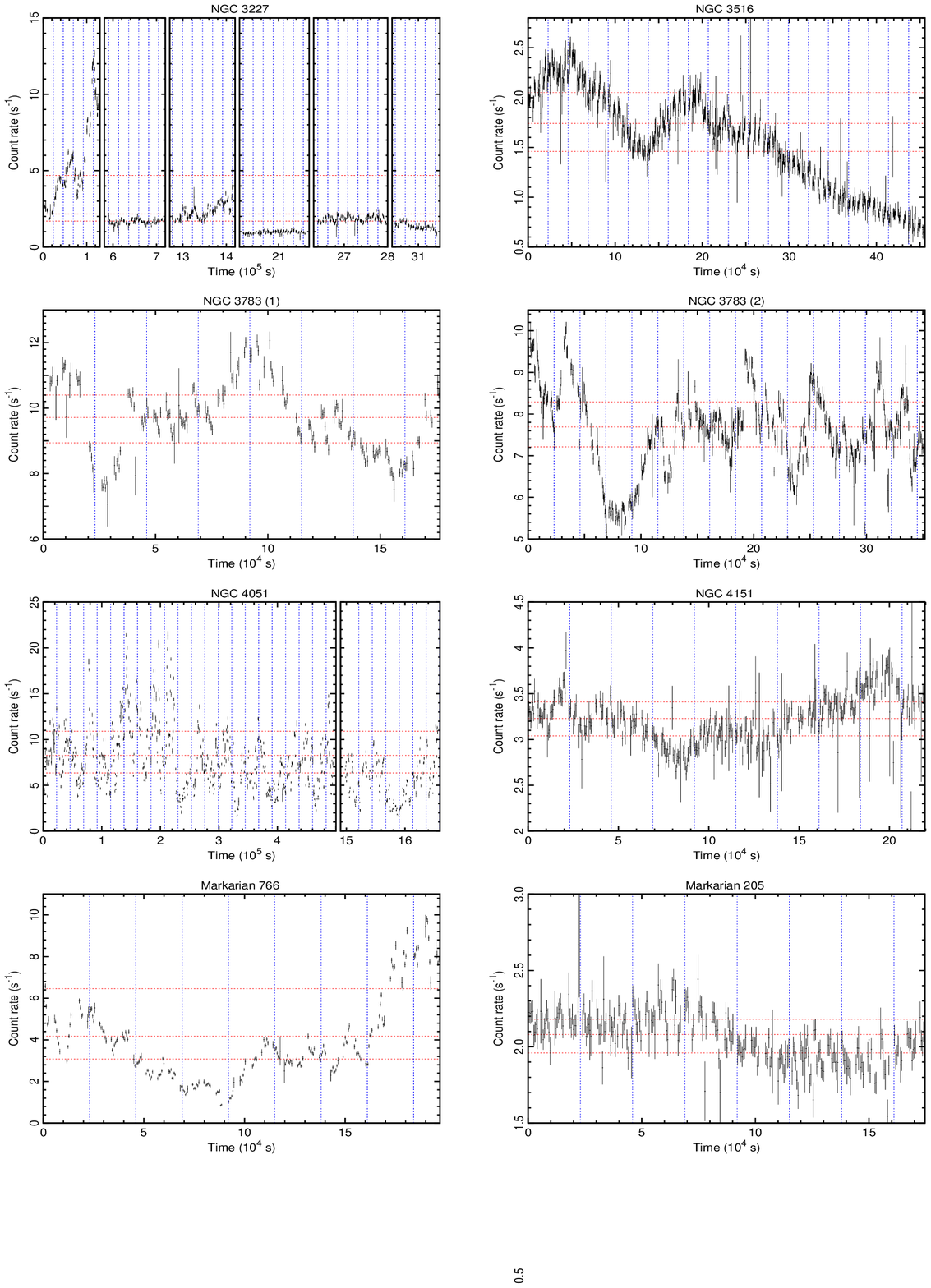}
\end{center}
\caption{--- Continued}
	\label{f1}
\end{figure*}
\addtocounter{figure}{-1}
\begin{figure*}
\begin{center}
\includegraphics[trim=3cm 4cm 0.5cm 4cm,clip,width=18cm]{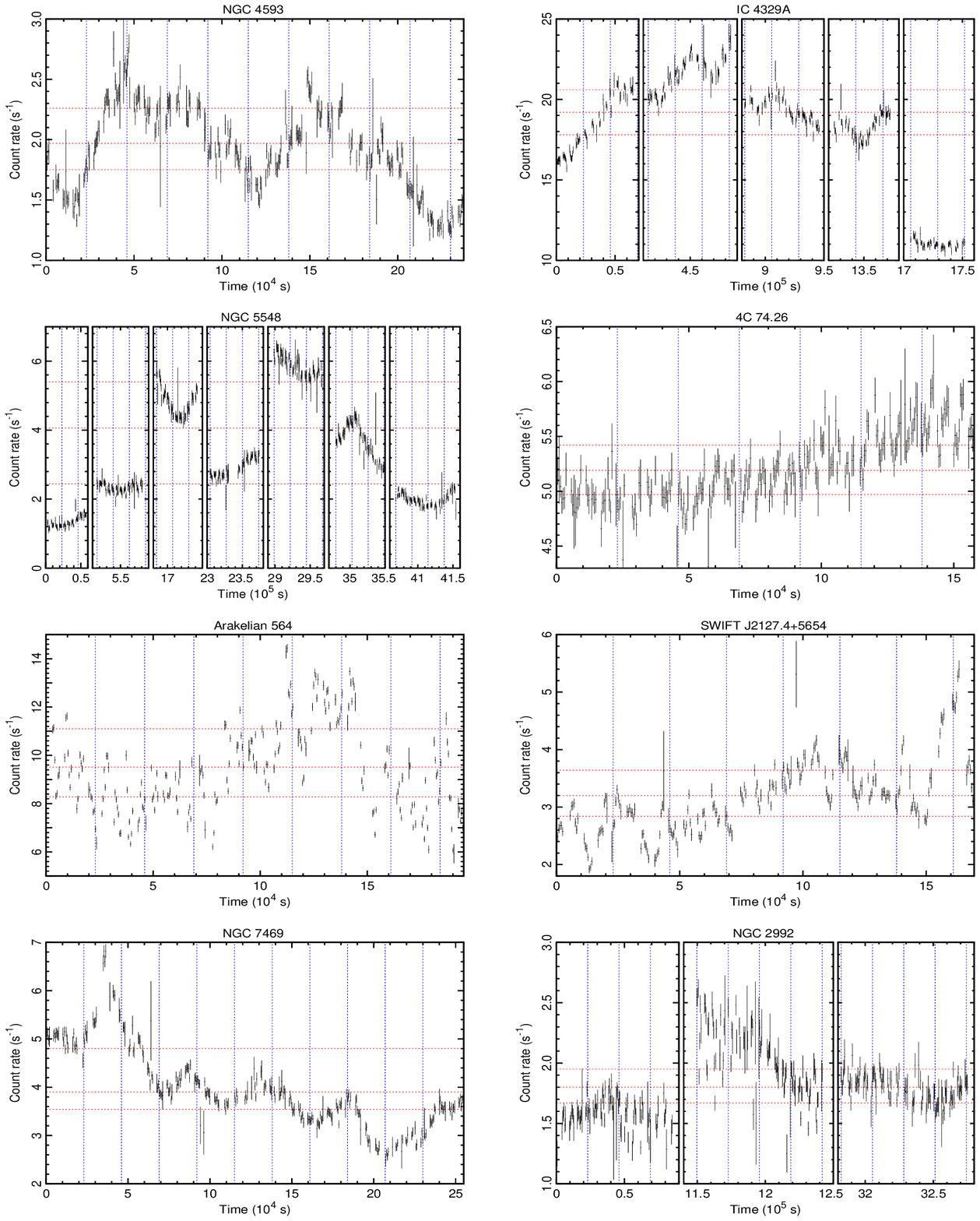}
\end{center}
\caption{--- Continued.}
\label{f1}
\end{figure*}
\addtocounter{figure}{-1}
\begin{figure*}
\begin{center}
\includegraphics[trim=3cm 4cm 0.5cm 4cm,clip,width=18cm]{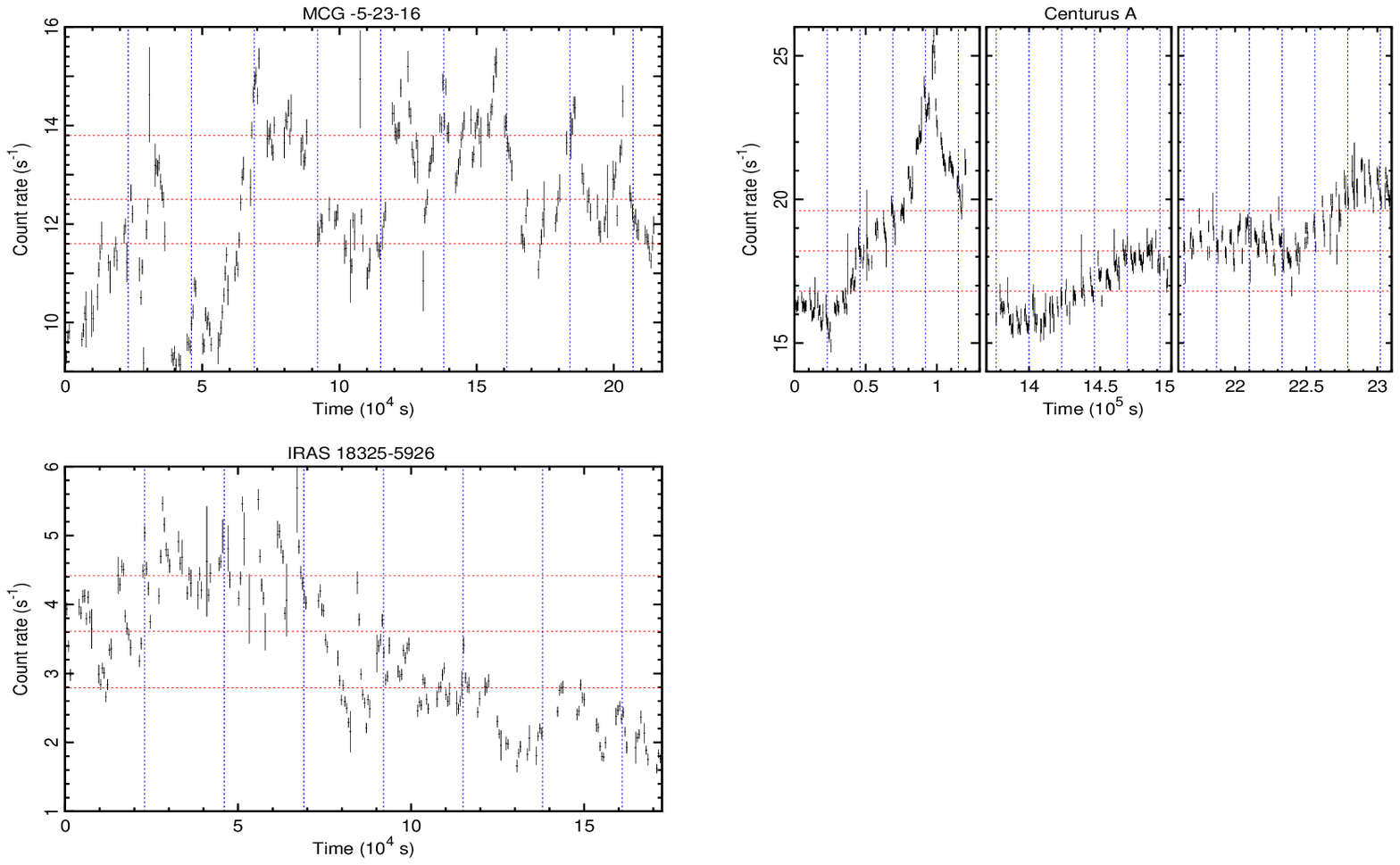}
\end{center}
\caption{--- Continued}
\label{f1}
\end{figure*}
%

\subsection{Time-averaged spectra}\label{ss3-1}
Since we are primarily interested in the iron K-line band, we use only 2 -- 10 keV of the 
XIS data for spectral fitting,  to avoid  complexities in the softer ($<$ 2 keV) energy band.
We also use the PIN data in 10 -- 40 keV, which help to constrain the underlying continuum
to study  iron K-band feature.

Firstly, we analyze the time-averaged spectra of  the 27 observations
from the 25 sources.
We apply  the same  ``3-component model'' proposed by  MEI2012.
The model is represented  as
\begin{equation}
F = W_{H} W_{L} \left( N_{1} + W_{2}N_{2}\right)P + RPN_{3} + I_{\rm Fe},
\label{eq1}
\end{equation}
where $P$ is the intrinsic cut-off power-law spectrum, 
$N_{1}$ and $N_{2}$ are the direct power-law normalization and the absorbed power-law normalization, respectively. $W_{H}$, $W_{L}$, and $W_{2}$ represent 
the transmissions due to 
high-ionized warm absorber, low-ionized warm absorber, and partial heavy absorber ($N_{\rm H} \gtrsim 10^{24}$~cm$^{-2}$), respectively. 
Each warm-absorber has two parameters, the  hydrogen column-density, $N_H$,  and the  ionization parameter, $\xi$, such that $W_{H}=\exp(-\sigma(\xi_H) N_{H, H})$,  
$W_{L}=\exp(-\sigma(\xi_L) N_{H, L})$ and $W_{2}=\exp(-\sigma(\xi_2) N_{H, 2})$,
where $\sigma(\xi)$ means the energy-dependent photo-absorption cross-section at $\xi$.
$R$ and $N_{3}$ are  the reflection albedo and the reflection normalization by the neutral accretion disk, respectively  (so that $RPN_3$ is the disk reflection component), and 
$I_{\rm Fe}$ is a narrow iron K$\alpha$ emission line. The interstellar extinction is also included in the model fitting, but not explicitly shown in Equation~(\ref{eq1}).

We used the X-ray spectral fitting package XSPEC version 12.7.0 for the spectral analysis. 
For the interstellar absorption and the disk reflection, we adopted \texttt{phabs} and \texttt{pexrav} \citep{Magdziarz1995} in XSPEC, respectively. 
Following MEI2012, the cut-off energy 
 and the disk inclination angle are fixed at  160 keV and $30^\circ$, respectively. 
$N_{3}$ is linked to  $N_{1}$ so that $N_{3}/N_{1}\sim\Omega/2\pi = 0.3$, where $\Omega$ is the solid-angle of the disk seen from the central source; 
we  confirmed that changing the inclination angle only slightly changes the best-fit spectral parameters within statistical errors.
For the warm absorbers, we use the table-grid model calculated 
by   MEI2012 using XSTAR (version 2.1kn8), where redshift was fixed at
0.001.

In the following, 
we set the  acceptance criterion of the successful model  fitting as the reduced $\chi^{2} < 1.2$.
The 4 sources  did not satisfy the criterion; NGC\,1365 (reduced $\chi^{2} =1.48$), NGC\,3227 (1.23), NGC\,4151 (1.24), and Centaurus\,A (1.42).
 3C\,111 satisfied the fitting criterion  (reduced $\chi^{2}=1.01$), but 
 the heavily absorbed component ($W_{2}N_{2}$) was not required.
Since the heavily absorbed component is the most important player in our  VPC model 
to produce the broad iron line feature and the spectral variation below $\sim$10 keV, 
we drop  3C\,111 from further study.

Consequently,  22 observations out of 20  sources 
are successfully represented with the  3-component model  (Figure~\ref{f2}), and used for further study. 
The best-fit parameters of the 22 time-averaged spectra  are summarized in Table~\ref{t2}.


In the case of MCG-6-30-15,  all the high-ionized  absorber ($W_{H}$), low-ionized absorber ($W_{L}$), and the heavy absorber ($W_{2}$) were required  
(MEI2012). However, we found the three absorbers are not always necessary.
In fact, only one of the 22 spectra requires all the three absorbers (we call it as Group A).
In addition to the heavy absorber,   7 spectra require only the high-ionized 
absorber (Group B), and one requires only the low-ionized absorber (Group C).  13 require none besides the heavy absorber (Group D). 
We accept these differences  as  diversity of the Seyfert galaxies, for the time being, being not able to 
find an obvious reason to explain the difference.  Our model for Group A, B, C and D are illustrated in Figure \ref{figABCD}.

\begin{figure*}
  \begin{center}
\includegraphics[trim=3cm 4cm 0.5cm 4cm,clip,width=18cm]{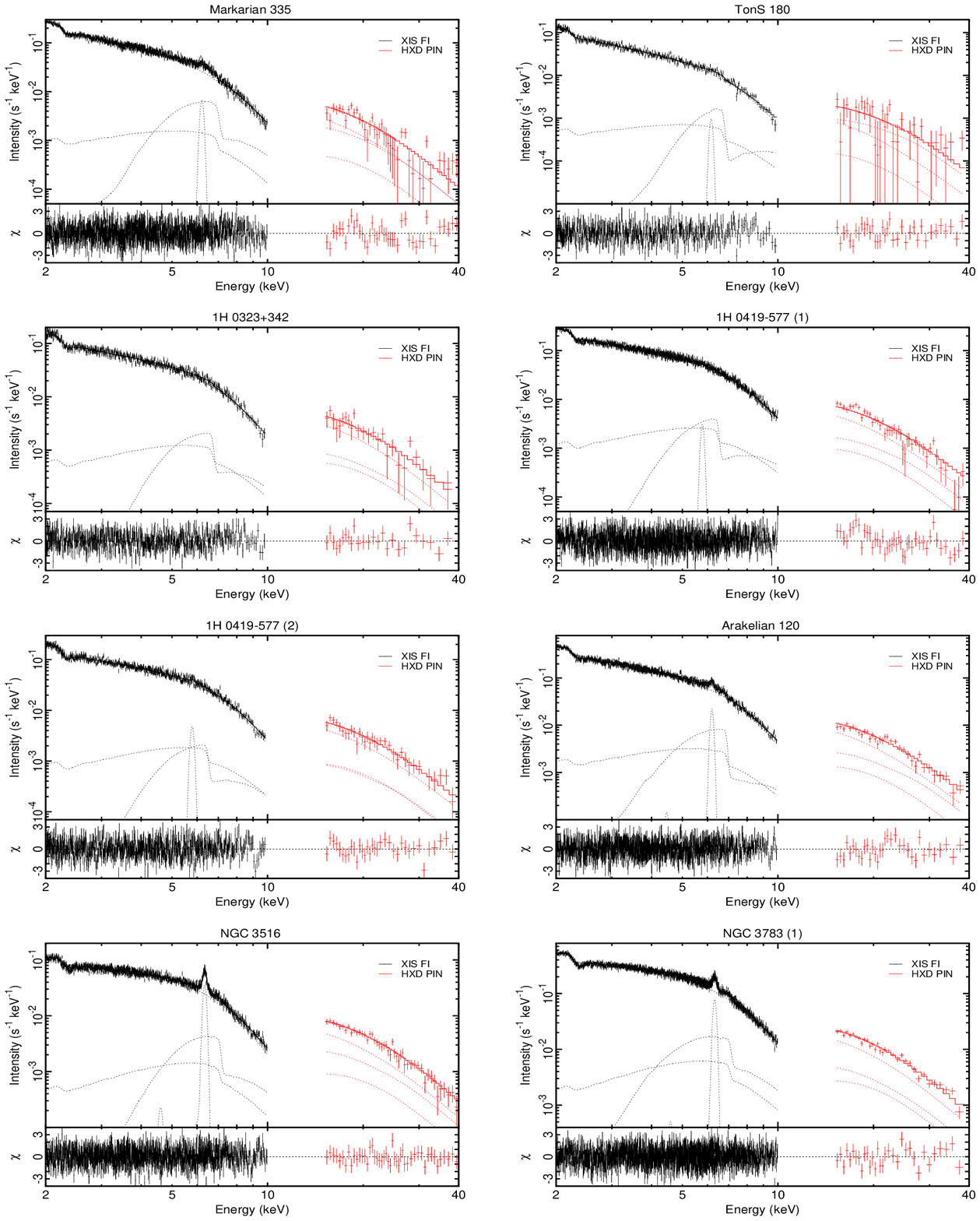}
  \end{center}
	\caption{Background-subtracted time-averaged XIS and PIN spectra. The data and the best-fit 
model are in the upper panel, while the residuals are in the lower panel.}
	\label{f2}
\end{figure*}
\addtocounter{figure}{-1}
\begin{figure*}
\begin{center}
\includegraphics[trim=3cm 4cm 0.5cm 4cm,clip,width=18cm]{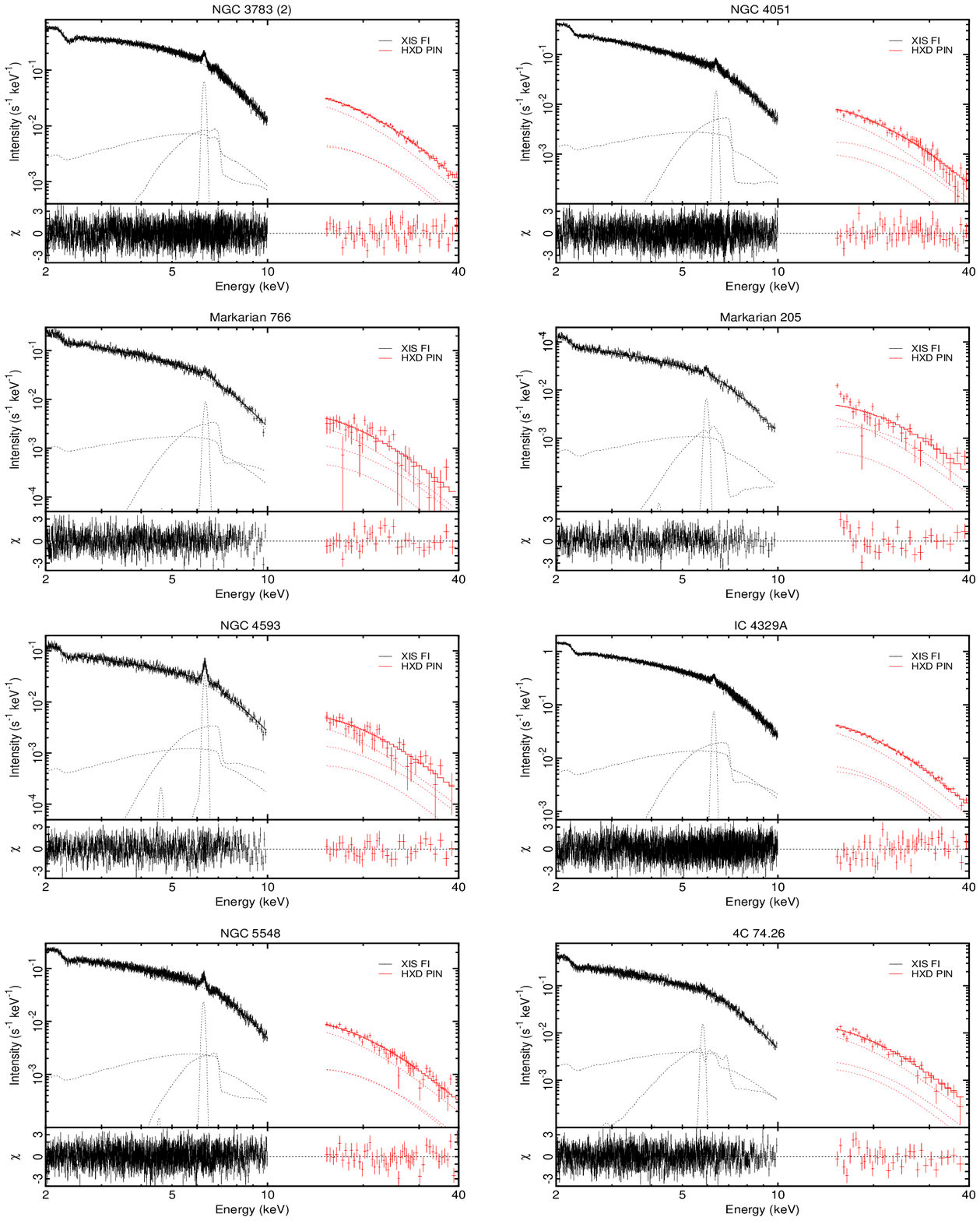}
\end{center}
\caption{--- Continued.}
\label{f2}
\end{figure*}
\addtocounter{figure}{-1}
\begin{figure*}
\begin{center}
\includegraphics[trim=3cm 4cm 0.5cm 4cm,clip,width=18cm]{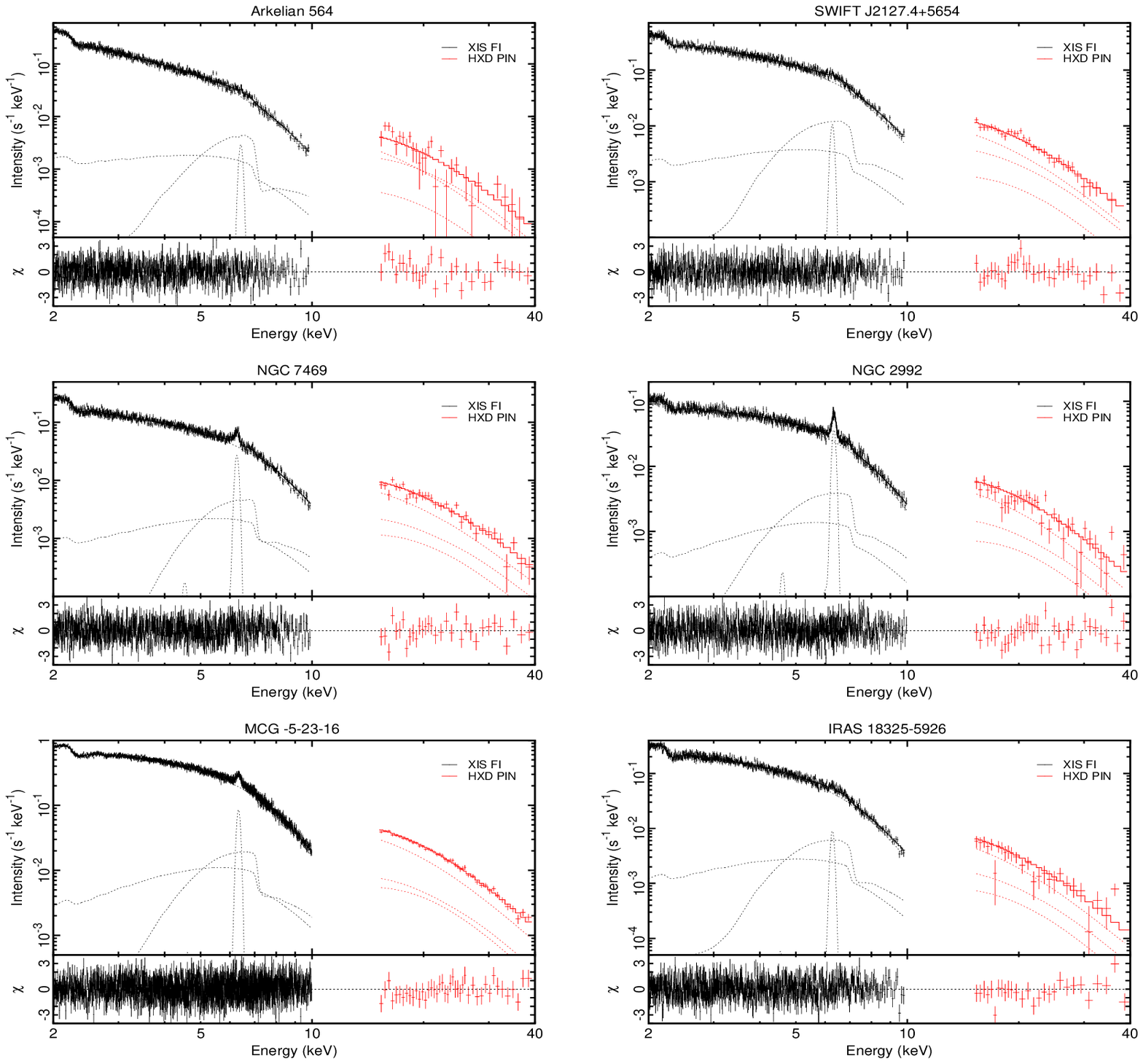}
\end{center}
	\caption{--- Continued.}
	\label{f2}
\end{figure*}

\begin{figure*}
  \begin{center}
\includegraphics[trim=2cm 9cm 0cm 3cm,clip,width=17cm]{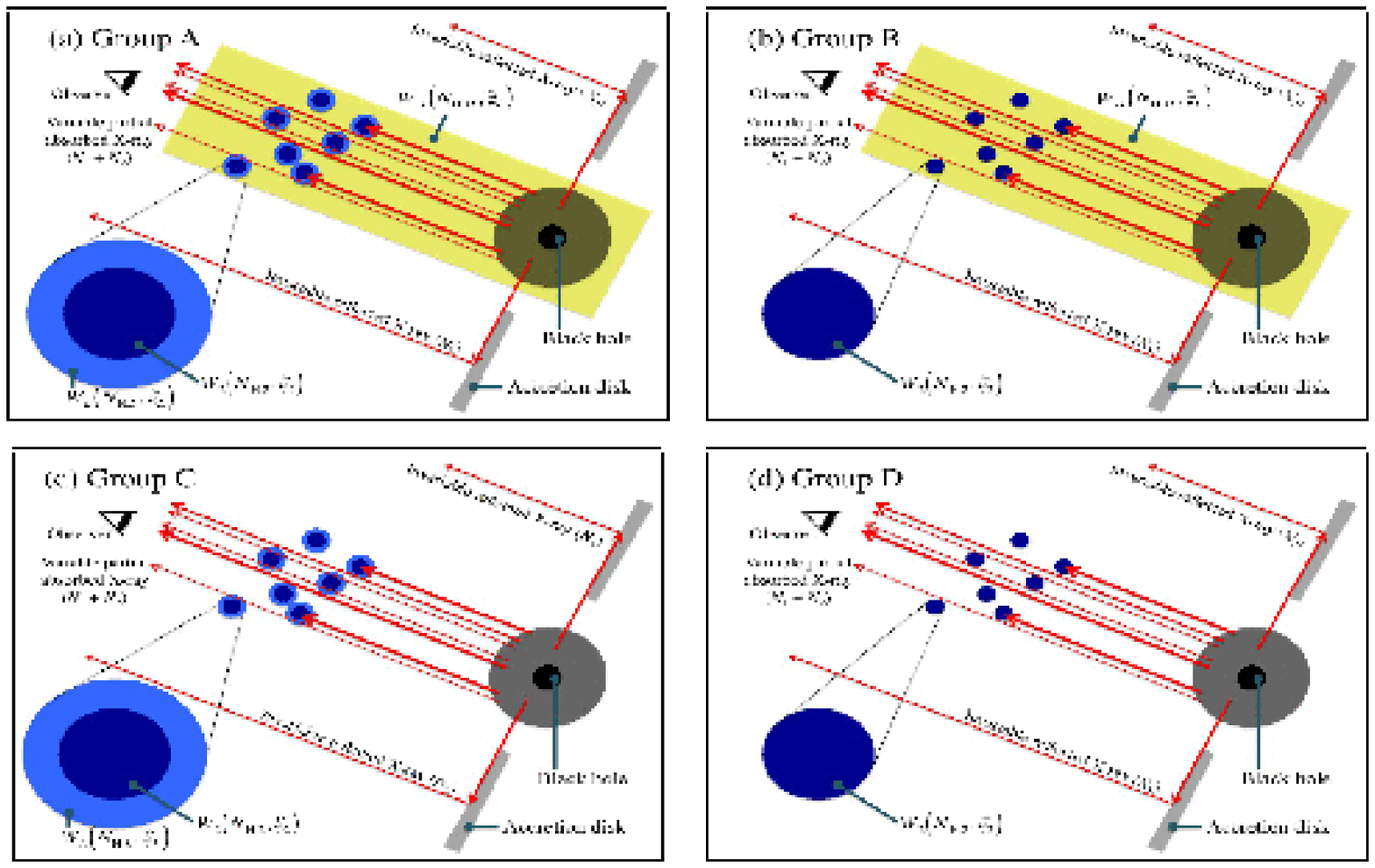}
  \end{center}
  \caption{A model for  four groups of the spectra we analyzed.  Group A requires all the three absorbers; 
the heavy partial absorber as the core of the absorbing clouds, the  low-ionized absorber as an  envelope of the 
clouds, and the uniformly surrounding high-ionized absorber.  Group B requires the partial heavy absorber
and the high-ionized absorber.  Group C requires the partial heavy absorber and the low-ionized absorber.  Group D requires only the partial heavy absorber.}
  \label{figABCD}
\end{figure*}

\subsection{Intensity-sliced spectra}\label{ss3-2}
Next, we study spectral variations during each of  the 22 observations
(out of 20 sources), the  average spectra of which were 
successfully fitted by the 3-component model.
We  examine if  the VPC model
proposed by MEI2012 may explain the observed spectral variations 
or not.
\subsubsection{Model}
To prepare,   we briefly review the VPC model in the following:
Let's define the ``total normalization'',  $N$, and the partial covering fraction, $\alpha$, 
so that
\begin{equation}
N = N_{1} + N_{2},
\label{eq2}
\end{equation}
and
\begin{equation}
N_{1} = (1 - \alpha) N,\;  N_{2} = \alpha N.
\label{eq3}
\end{equation}
Here, we consider  such a situation that the total normalization, $N$,  represents the
intrinsic AGN luminosity, and the  X-ray source having a finite size is partially
covered by fragmented heavy absorbing clouds (with the absorption $W_2$),  where the
geometrical partial  covering fraction is  $\alpha$.

Detailed study of 
spectral variations of MCG-6-30-15 by MEI2012 revealed, surprisingly,  
that  the low-ionized warm absorption $W_L=\exp(-\sigma(\xi_L) N_{H,L})$ is 
related to  the same partial covering fraction $\alpha$, such that 

\begin{equation}
N_{{\rm H},L} = \alpha <N_{{\rm H},L}>,
\label{eq4}
\end{equation}
where $<N_{H,L}>$ is the common amount  of the  column density of the
low-ionized absorber while the spectra vary.  This relation was  unexpected, since the
low-ionized absorption $W_L$ and the partial covering fraction due to $W_2$ should be  in principle independent (see equation\ \ref{eq1}).
MEI2012 interpreted this relation as that 
the low-ionized absorber ($W_L$) and the heavy absorber ($W_2$)
are parts of the same absorbing clouds (Figure \ref{figABCD}), so  that the 
``double partial covering'' with the  same covering fraction is presumably the case (MES2014).

Consequently, the 3-component model (\ref{eq1}) is rewritten as
$$
F= \exp\left(-\sigma(\xi_H)N_{H,H}\right) \exp\left(-\sigma(\xi_L)\, \alpha\,  <N_{H,L}>\right) 
$$
$$
\left( 1 - \alpha + \alpha \exp\left(-\sigma(\xi_2)N_{H,2}\right) \right)PN
$$
\begin{equation}
~~~~~~~~~~~~~~~~~ + RPN_{3} + I_{\rm Fe}. 
\label{eq8}
\end{equation}
MEI2012 applied this model to the eight  intensity-sliced spectra
of MCG-6-30-15 observed with Suzaku, where the good-time intervals (GTIs) are determined  using XIS below 10 keV.
$N_3$ is assumed to be invariable, and fixed to be 0.3 times the average of $N_1=(1 - \alpha) N $.
The iron line parameters are also fixed to those obtained from the time-averaged spectra. 
These  spectra were explained well in 1 -- 40 keV by  only variation of $\alpha$, whereas $N$ is invariable  besides the  brightest
one (where $N$ is 1.5 times greater);
all the other spectral  parameters are invariable (=common to all the variable spectra). 

\subsubsection{Model Fitting}
Following MEI2012, we create the intensity-sliced spectra based on the XIS flux, and
examine the VPC model (\ref{eq8}).
For each observation, we define four intensity ranges in the 0.2 -- 12.0~keV light curve 
so that the photon counts for each intensity range be approximately equal, 
and create the four intensity-sliced energy spectra. The red dotted horizontal lines in Figure~\ref{f1} indicate the intensity boundaries for each observation.

We try to fit the four intensity-sliced spectra in 2 -- 40 keV simultaneously with the model (\ref{eq8}) only making $\alpha$ variable.
If necessary,  $N$ is varied in addition. Other parameters are common to 
all the four spectra, and the best-fit values are determined from the model fitting (Table \ref{t3}).

As a result, we successfully fitted the intensity-sliced spectra of the 20 observations of 18 sources 
only varying $\alpha$. The rest two sources (Markarian\,766 and NGC\,5548) require 
$N$ to be slightly variable:
Markarian\,766 requires $N$ for the brightest spectrum to be $\sim$1.3 times
greater than the rest.
NGC\,5548 requires the  $N$ for the brighter two spectra to be $\sim$1.7 times
higher than the dimmer two.
 We summarize the spectral fitting results with the VPC model
for the 22 intensity-sliced spectra in 
Figure~\ref{f3} and Table~\ref{t3}.
\begin{figure*}[t]
\begin{center}
\includegraphics[trim=3cm 4cm 0.5cm 4cm,clip,width=18cm]{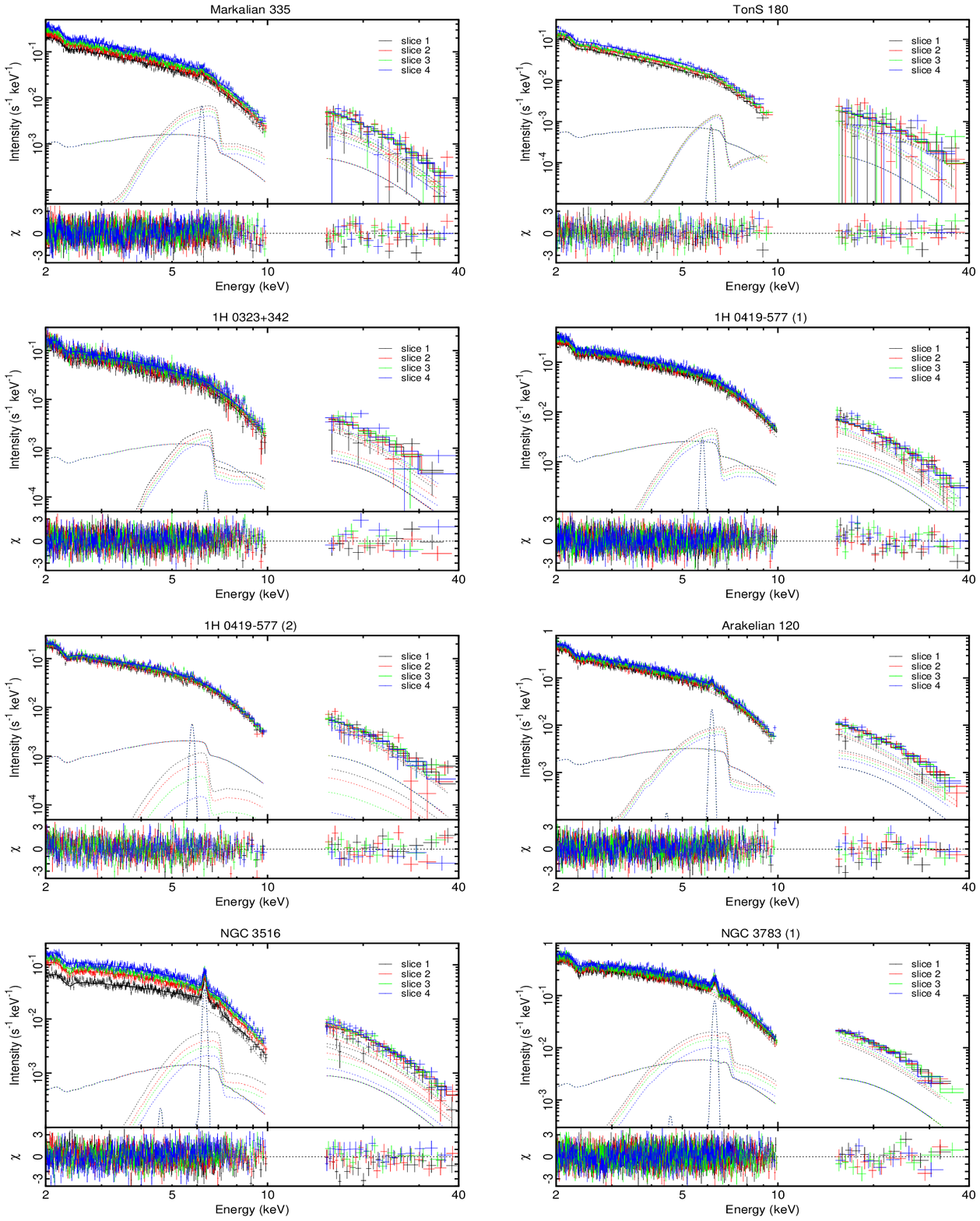}
\end{center}
\caption{Model fitting results of the intensity-sliced XIS and PIN spectra.
For each observation, the four intensity-sliced spectra are made using XIS with approximately equal counts for each intensity range, and 
the same GTIs are used to extract  PIN spectra.
The four spectra are fitted simultaneously only varying the partial covering fraction, $\alpha$,
except Markarian 766 and NGC 5548 where $N$ is also  slightly varied (see texts).
The data and the best-fit power-law model are in the upper panel, while the residuals are in the lower panel.}
	\label{f3}
\end{figure*}
\addtocounter{figure}{-1}
\begin{figure*}[t]
\begin{center}
\includegraphics[trim=3cm 4cm 0.5cm 4cm,clip,width=18cm]{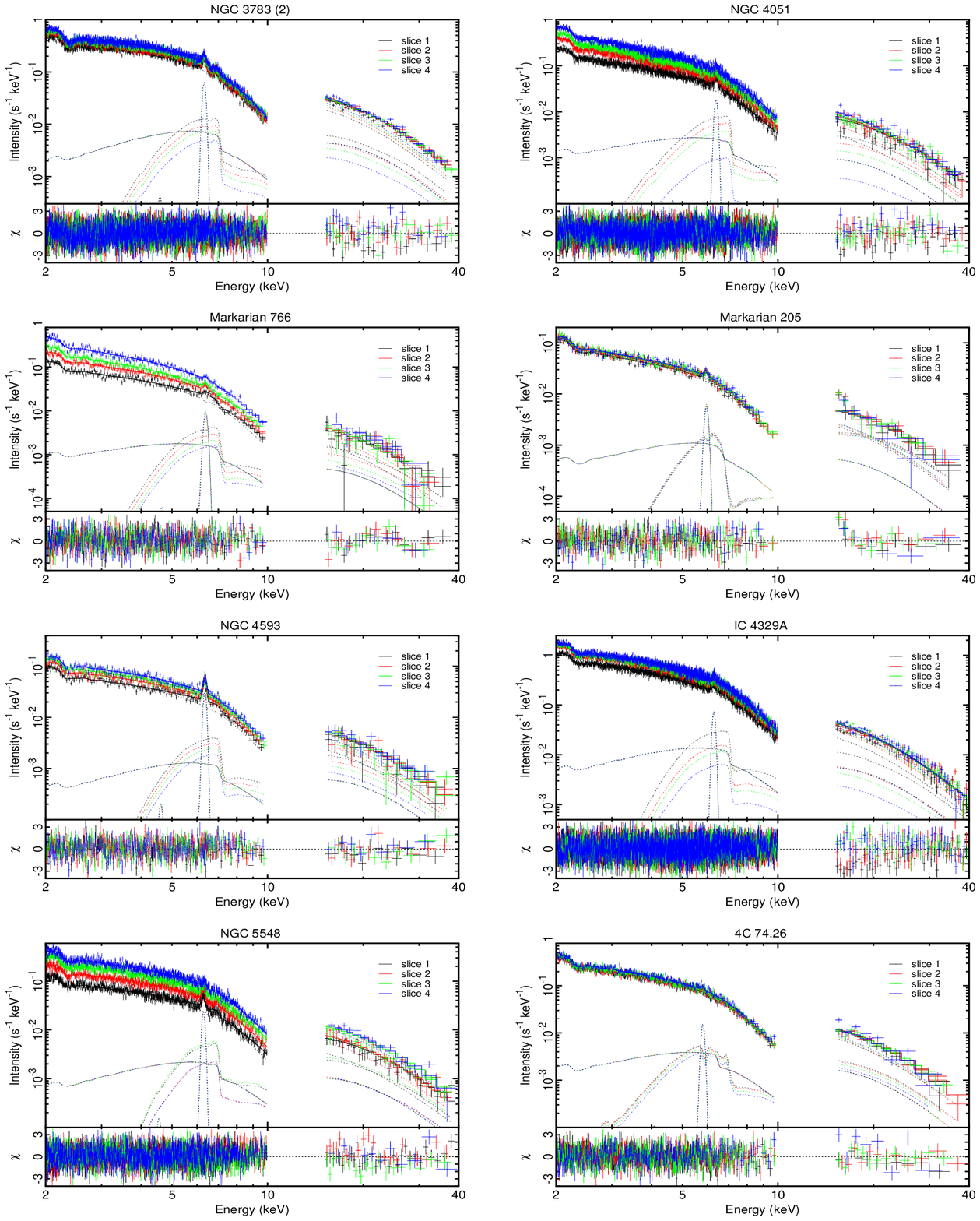}
\end{center}
	\caption{--- Continued.}
	\label{f3}
\end{figure*}
\addtocounter{figure}{-1}
\begin{figure*}[t]
\begin{center}
\includegraphics[trim=3cm 4cm 0.5cm 4cm,clip,width=18cm]{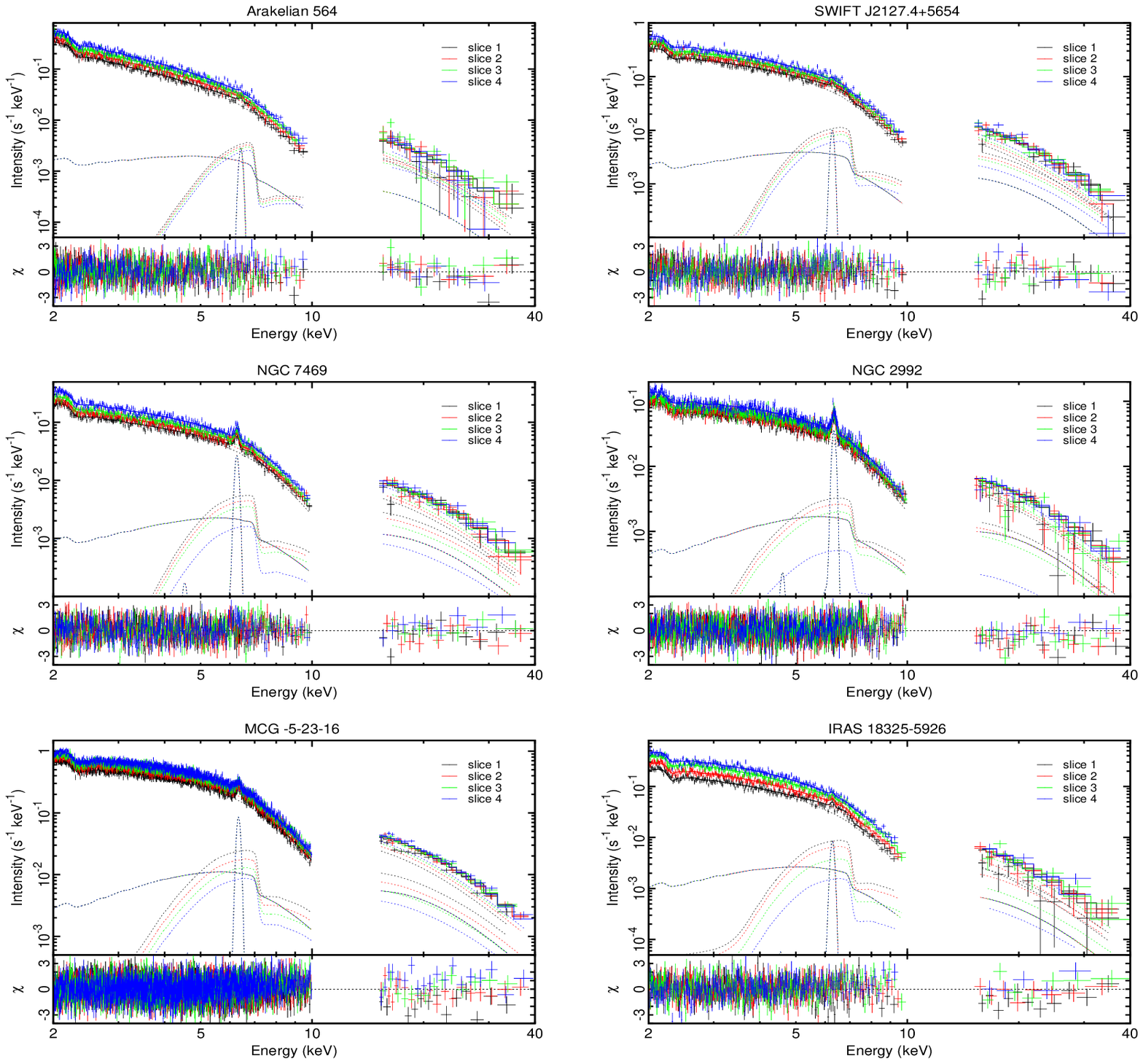}
\end{center}
	\caption{--- Continued.}
	\label{f3}
\end{figure*}

\subsection{Time-sliced spectra}\label{ss3-3}
Next, we see if  time sequences  of the soft X-ray energy spectra 
are also  explained by the VPC model with the time-varying partial covering fractions.
In the XIS light-curves, we define  boundaries  of the
time-sliced spectra  every $2.3\times10^{4}$~sec, which corresponds to four Suzaku orbital periods. The blue 
vertical  dotted lines in Figure~\ref{f1} show the time-boundaries for each
 observation, and we made energy spectra from
each time bin. 
Numbers  of the time-sliced spectra  in a single observation are from  6 (4C 74.26) to 28 (NGC 4051).
For each observation, a time-series  of the  XIS and PIN spectra
are fitted simultaneously with the VPC model  in Equation (\ref{eq8}), while spectral variation above 10 keV is hardly
constrained, because the photon statistic is not sufficient.
We first try to fit the spectra only allowing the partial covering fraction $\alpha$ to vary.
If not successful (i.e., $\chi^2_r > 1.2$), the total normalization $N$ is varied in addition.

As a result, we found that the XIS spectral variations  are explained only varying $\alpha$ for 21 observations from  19 sources.
Figure~\ref{f4} shows  variation of the observed counting rates in 0.2 -- 12 keV
  and the partial covering fraction 
for each of the 21 observations. Anti-correlation between
the counting rate and the partial covering fraction is obvious. Namely, 
apparent soft X-ray flux and spectral variations  below $\sim$10 keV
are primarily caused by variation of the partial covering 
fraction.
Only the remaining source, NGC\,5548, needs varying the total normalization $N$ in addition to $\alpha$
within a single observation.   Figure~\ref{f5} shows variation of the counting rates and
partial covering fraction for  NGC\,5548.  Note that observation of NGC5548 spans more than 
$4 \times 10^6$ sec, which is the longest in our samples, and 
the total normalization $N$ is still invariable 
within time-scales less than several  $  10^5$ sec.
This suggests that the intrinsic luminosity variation of
Seyfert galaxies has a timescale longer than $\sim10^5$
sec.

\subsection{Variation in the Iron Line Energy Band}\label{ss3-4}
In order to study variation in the iron line energy band, 
we calculate the Root Mean Square (RMS) spectra   for the 22 observations.
We use the same time-series of the spectra used in the previous section.
Namely, RMS spectra are calculated with a bin-width of $2.3\times10^{4}$~sec.

  For a time-series of $\left\{x_{i}\pm\delta x_{i}\right\}_{i=1,N}$, where $\left\{x_{i}\right\}$ are 
background-subtracted counting rates, $\left\{ \delta x_{i}\right\}$ are their errors, and $N$ is number of the time-bins, the RMS variability is given as
\begin{equation}
\mathrm{RMS\, \,  variability} = F_{\rm var}=\frac{\sqrt{\frac{1}{N-1}\sum_{i=1}^N(x_i-\bar{x})^2-\frac{1}{N}\sum_{i=1}^N\delta x_i^2}}{\bar{x}}
\label{eq9}
\end{equation}
where ${\bar x}= (\sum_{i=1}^N x_i)/N$, and the error is
\begin{equation}
{\rm RMS \, \, error} = \frac{1}{F_{\rm var}}\sqrt{\frac{1}{2N}}\; \frac{\frac{1}{N-1}\sum_{i=1}^N(x_i-\bar{x})^2}{\bar{x}^2}
\end{equation}
 (Edelson  et al.\ 2002). 
 For each observation, we computed the RMS variability
for 15 energy band with Equation~(\ref{eq9}) to constitute the RMS spectrum. We used only the XIS data in 2 -- 10~keV to focus the iron line energy band.
The 22 RMS spectra are shown in Figure~\ref{f6} (in black). 

Since each time-sliced spectrum is fitted successfully with the VPC model, the RMS model spectra can
be calculated from the best-fit VPC model spectra.  These model RMS spectra are also shown in 
Figure~\ref{f6} (in red). 
We find that  the observed RMS spectra and the model RMS spectra  agree well, and both tend to  show  drops at the
iron line energy band (6 -- 7 keV).  The reason for that 
is discussed in section \ref{rmsdiscussion}.

\begin{figure*}[t]
\begin{center}
\includegraphics[trim=3cm 4cm 0.5cm 4cm,clip,width=18cm]{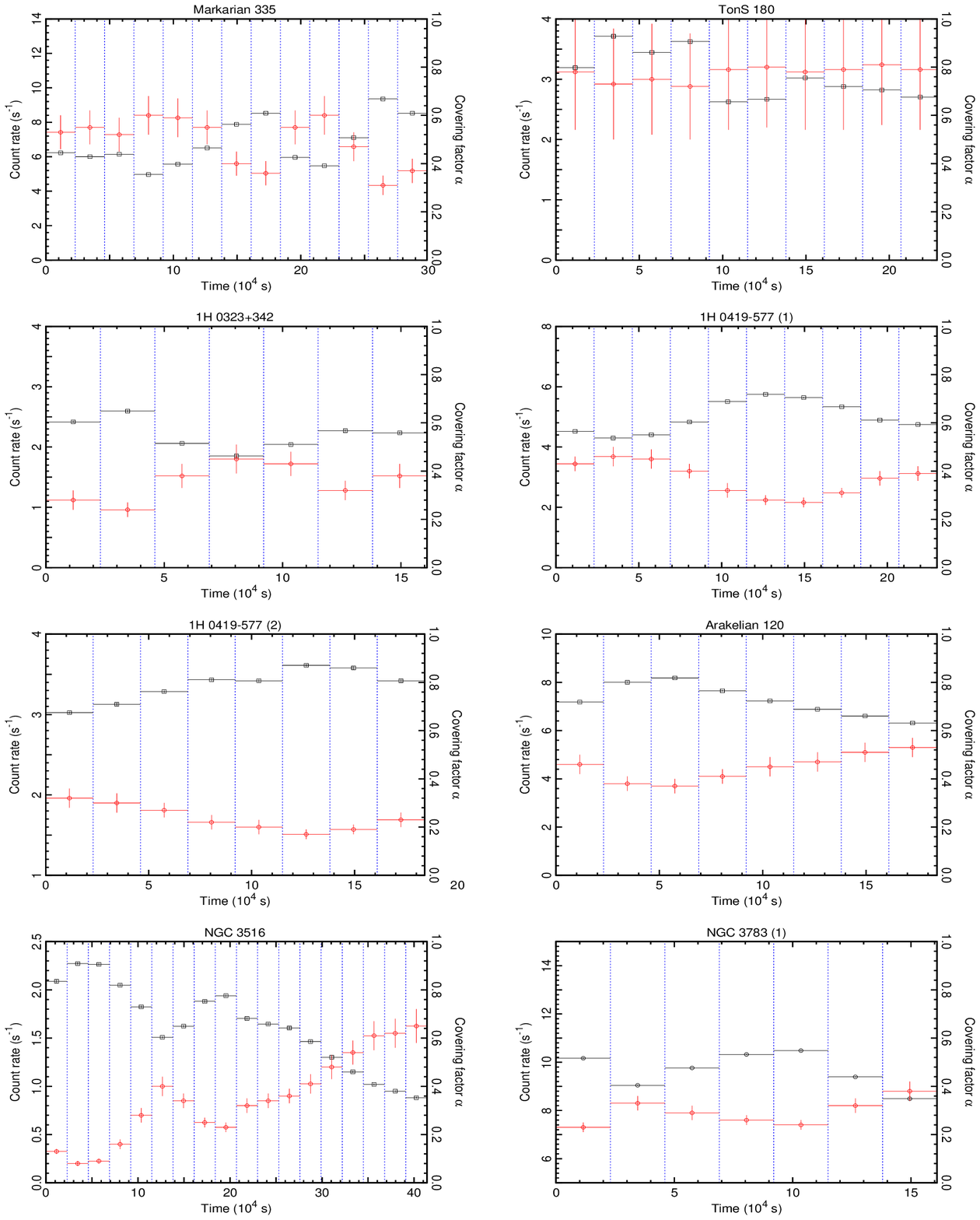}
\end{center}
	\caption{Variations of the observed XIS counting rates (0.2 -- 12 keV;
black, scale in left)
and the partial covering fractions (red, scale in right) for 21 observations.}
	\label{f4}
\end{figure*}
\addtocounter{figure}{-1}
\begin{figure*}[t]
\begin{center}
\includegraphics[trim=3cm 4cm 0.5cm 4cm,clip,width=18cm]{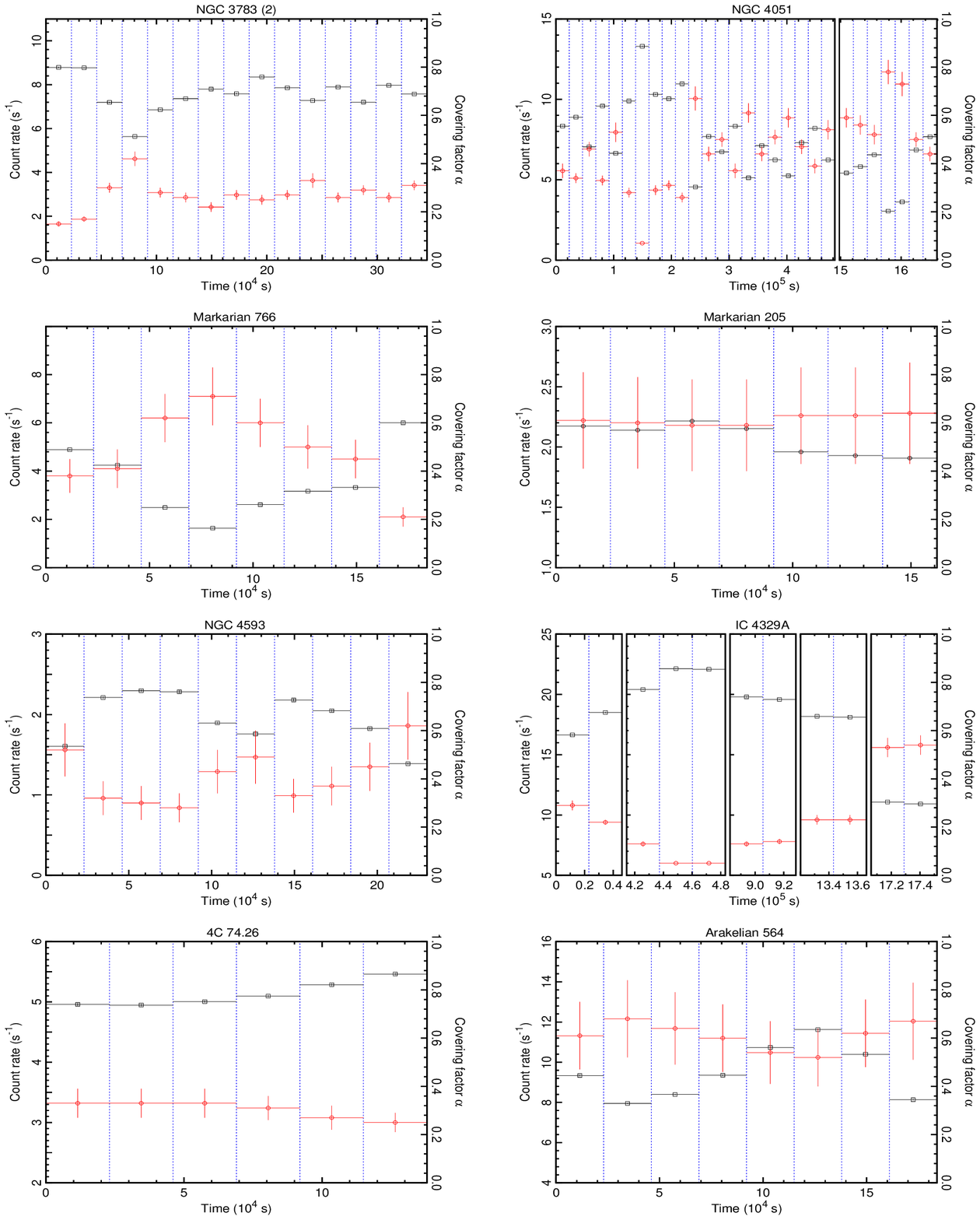}
\end{center}
\caption{--- Continued.}
\label{f4}
\end{figure*}
\addtocounter{figure}{-1}
\begin{figure*}[t]
\begin{center}
\includegraphics[trim=3cm 4cm 0.5cm 4cm,clip,width=18cm]{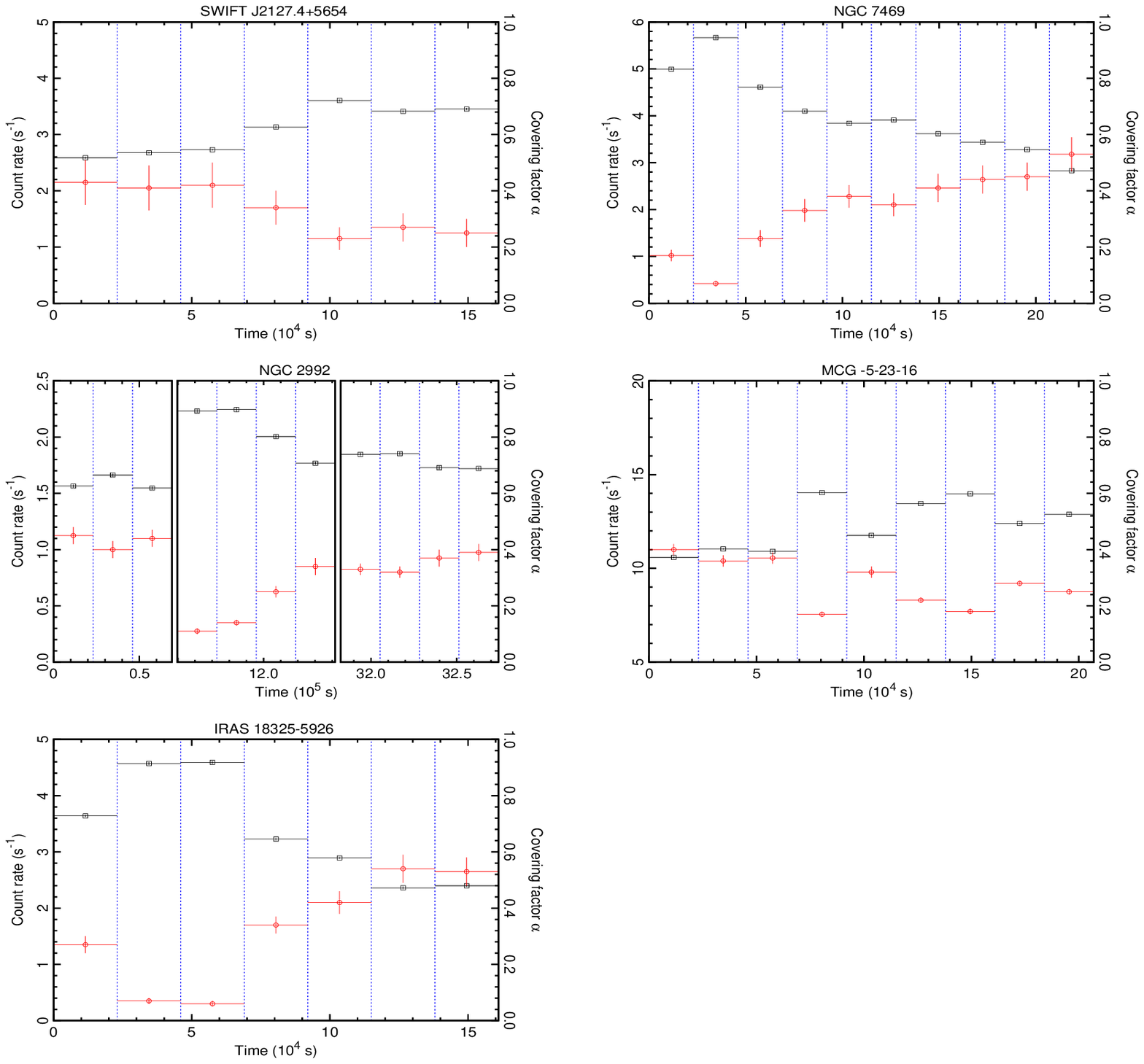}
\end{center}
\caption{--- Continued.}
\label{f4}
\end{figure*}
%

\begin{figure*}[t]
	\begin{center}
          \includegraphics[width=120truemm,height=80truemm]{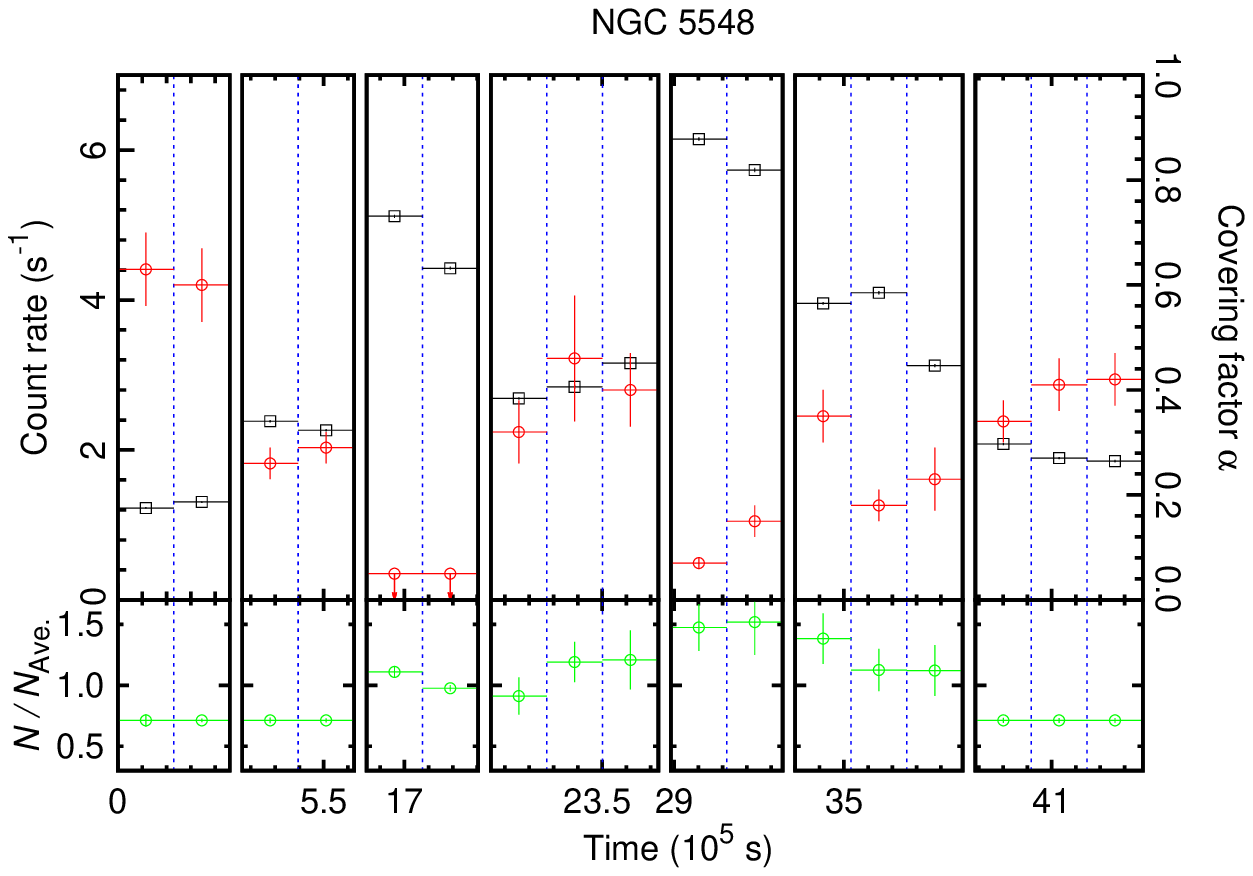}
	\end{center}
	\caption{Top: Variation of the observed XIS counting rates (0.2 -- 12 keV;
black, scale in left) and the partial covering fractions (red, scale in right) for NGC5548. Bottom: Ratio of the total normalizations to the
average, indicating variation of the the intrinsic luminosity.}
	\label{f5}
\end{figure*}
%

\begin{figure*}[t]
\begin{center}
\includegraphics[trim=3cm 4cm 0.5cm 4cm,clip,width=18cm]{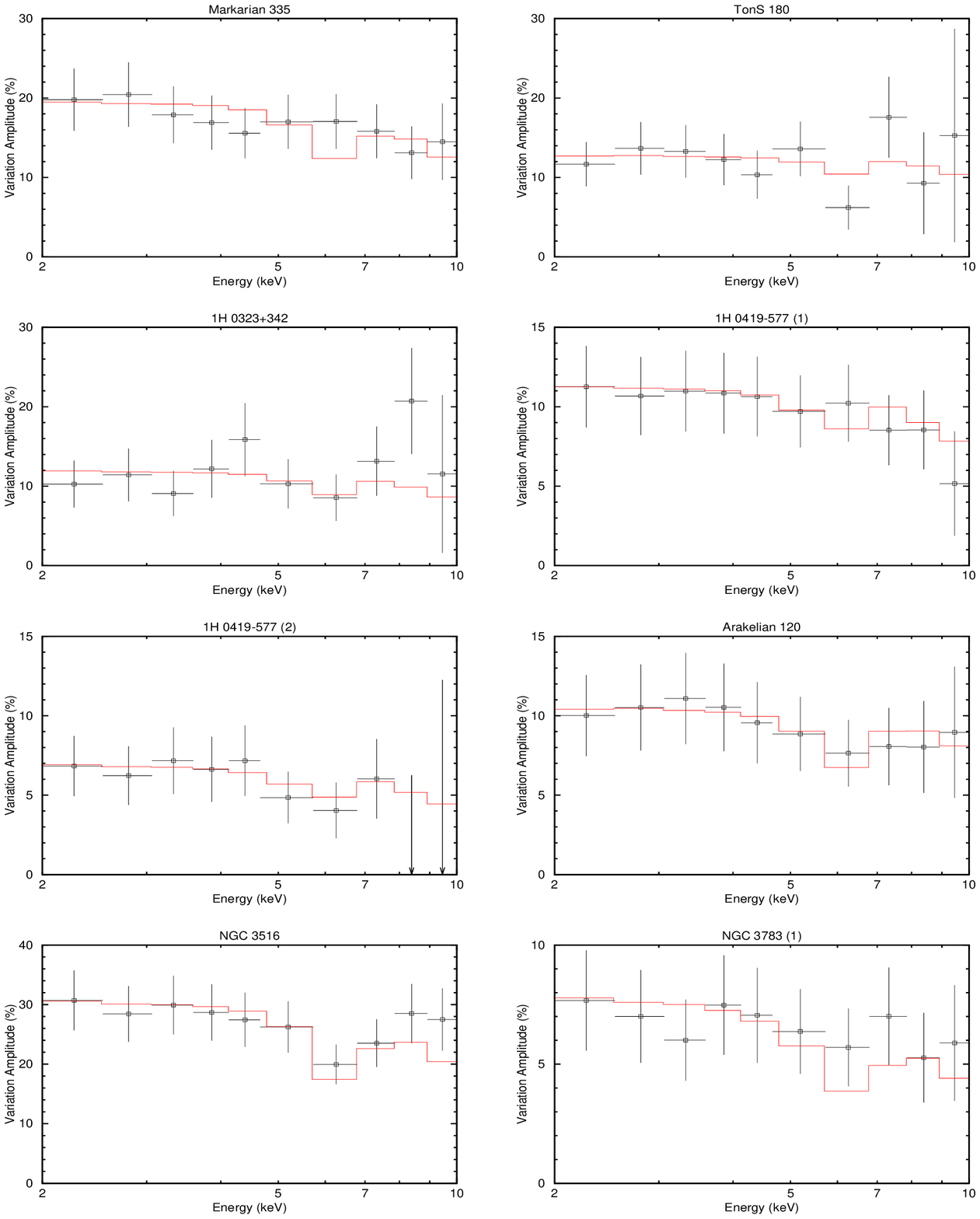}
\end{center}
	\caption{RMS (Root-Mean-Square) spectra for the 22 observations.  Black points are calculated from the data, and
                 red-histograms are calculated from the best-fit spectral models for the time-sequence spectra.}
	\label{f6}
\end{figure*}
\addtocounter{figure}{-1}
\begin{figure*}[t]
\begin{center}
\includegraphics[trim=3cm 4cm 0.5cm 4cm,clip,width=18cm]{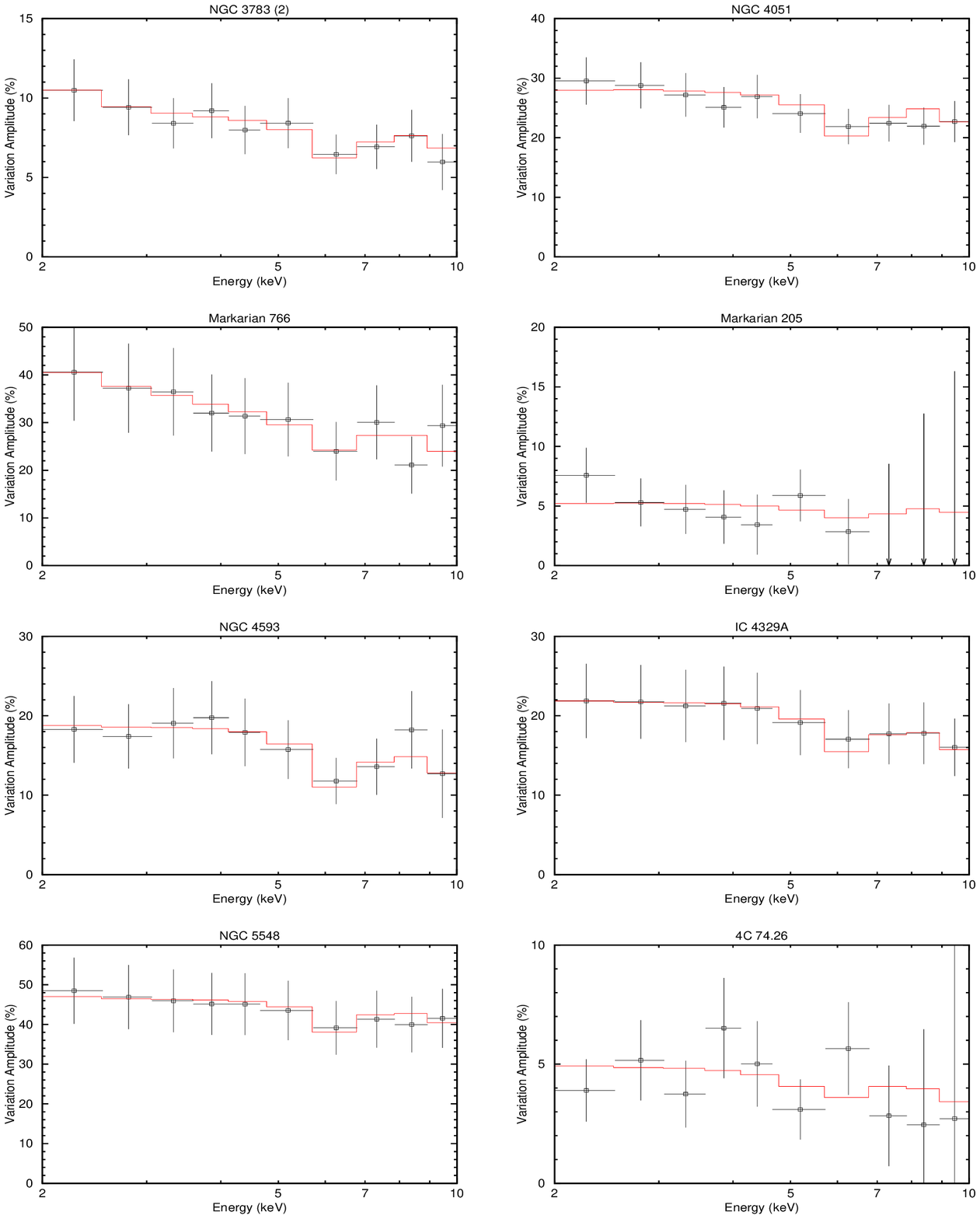}
\end{center}
	\caption{--- Continued.}
	\label{f6}
\end{figure*}
\addtocounter{figure}{-1}
\begin{figure*}[t]
\begin{center}
\includegraphics[trim=3cm 4cm 0.5cm 4cm,clip,width=18cm]{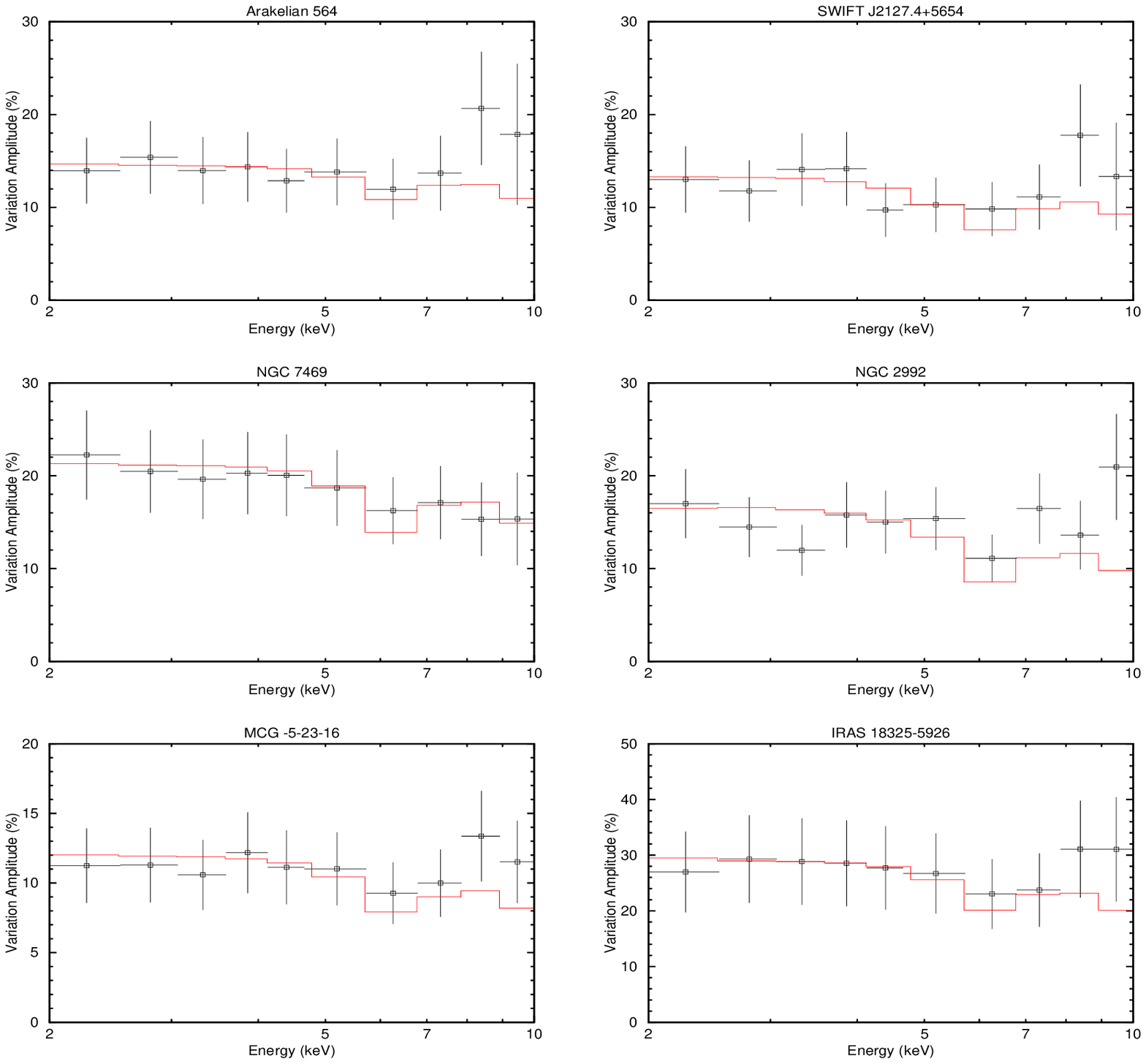}
\end{center}
	\caption{--- Continued.}
	\label{f6}
\end{figure*}
%

\section{Discussion}\label{s4}
\subsection{Origin of the Broad Iron Line Feature and the Soft X-ray Variation}\label{rmsdiscussion}
We have seen that the Variable Partial Covering (VPC) model, which has been 
proposed for MCG-6-30-15 (MEI2012) and  1H0707--495 (MES2014), 
can explain the 2--10 keV spectral variation of   20 other Seyfert galaxies observed with Suzaku.
The original VPC model for MCG-6-30-15 requires three types of the absorbers; optically-thick partial absorbing 
clouds, optically-thin envelopes of the clouds,  and a uniformly surrounding highly ionized absorber  (Group A in Figure \ref{figABCD}).
1H0707--495 does not require the highly ionized absorber (Group C in Figure \ref{figABCD}; MES2014), and  the
20 targets studied in this paper are grouped into four  depending on the number and types of the absorbers required   (Figure \ref{figABCD}).  The optically-thick
partial absorbers are always required,  where the hydrogen-column densities are  $\gtrsim10^{24}$ cm$^{-2}$ and the partial covering fraction is variable
from $\sim$0.1 to   $\sim0.7$ (Table \ref{t3}).

The heavily absorbed spectral components by such optically-thick absorbing clouds exhibit strong iron K-edge, which, together with the distant reflection feature (narrow iron emission line and weak iron K-edge), 
can explain the observed spectral feature in the iron K-band (Figure \ref{f2}).  Because such thick absorbers are completely  opaque at $\sim$2 keV (Figure \ref{f2}), variation of the
partial covering fraction can cause significant soft X-ray flux variations even if the intrinsic X-ray luminosity is invariable.
In fact, the observed flux and spectral variations below $\sim10$ keV within a time scale of $\lesssim$ day are explained by only
variation of the partial covering fraction, while the intrinsic luminosity below $\sim$10 keV is assumed to be constant  (Figures \ref{f3} and \ref{f4}).

It is  well-known  that  fractional variation at the broad iron line-like feature  in MCG-6-30-15 is significantly
reduced (\cite{Fabian2002}, \cite{Matsumoto2003}, \cite{Inoue2003}).
In the VPC model when the total normalization (intrinsic luminosity) is invariable, 
change of the partial covering fraction causes {\em anti-correlation}\/ between the direct (uncovered) spectral
component and the absorbed (covered) spectral component  below $\sim$10 keV  (Figure \ref{f3}). 
Thus, flux variations of the direct component and the absorbed  component {\em cancel each other}.
The fluxes from these  components are  the closest where the absorbing material is most transparent, 
just below the iron edge at $\sim$7.1 keV.   
Consequently,   the cancellation of the two spectral components works most
effectively at around $\sim$5.5--7 keV, and the fractional variation is reduced   (Figure \ref{f6}; see also 
 \cite{IME} and   MEI2012). 
Also, the $\sim$6.4 keV  iron emission line is not expect to vary with the continuum,  as it is emitted from  outer parts of 
the accretion disk.  These effects work together to reduce the fractional variation  at around  the iron energy band.

Among the datasets used in this paper, span of the NGC5548 observation is the longest, for $\sim$50 days. 
Its  soft X-ray luminosity {\em does}\/ vary, by a factor of two over $\sim10^6$ sec, while the covering 
fraction varies more significantly at much shorter timescales  (Figure \ref{f5}). Presumably, 
soft X-ray flux variation of Seyfert galaxies have two different origins with different timescales;  intrinsic luminosity variation over 
$\sim$days and  variation of the  partial covering fraction below  a timescale of $\sim$day (see also MES2014).



\subsection{Hard X-ray Variations}
We could fit the variable spectra in 2 -- 10 keV  only varying the partial covering fraction (Figures \ref{f3} and \ref{f4}).
When the same GTIs determined based on the XIS counting rates below 10 keV are used for PIN in 10 -- 40 keV, we see some, but not
very significant,  residuals above  $\sim$10 keV (Figure \ref{f3}); this  was also the case for MCG-6-30-15 (MEI2012).
Note, in the VPC model, variation of $\alpha$ hardly affects  the 
spectral variation  above $\sim$10 keV, where effect of the photo-absorption is minimum. 
Namely, if the hard-tail variations above $\sim$10 keV were correlated with the intensities below $\sim$10 keV,
we would have seen much more significant systematic discrepancies  above $\sim$10keV. Thus, the current results 
 suggest that the hard-tails above $\sim$10 keV are  rather independently variable and 
averaged when sorted by the intensities  below $\sim$ 10 keV.  In fact, presence of such  independently variable hard components 
above $\sim$10 keV is confirmed through  completely different approach in MCG-6-30-15 (Noda et al.\ 2011)
and NGC3516 (Noda et al.\ 2013).  For 1H0707--495, the VPC model almost perfectly represents the flux variation in 
0.5 -- 1 keV in timescales within $\sim$day, while less  satisfactorily in higher energies (MES2014); this  also suggests the 
commonness  of the independently variable hard component in Seyfert galaxies.
In summary, we propose that  observed X-ray flux/spectral variation of Seyfert galaxies is explained by variation of the
partial covering fraction in timescales below $\sim$day and the  intrinsic soft X-ray luminosity variation over
$\sim10^6$ sec, both of which are mainly responsible for  $\lesssim$10 keV, and independent hard X-ray variations above $\sim$10 keV.

\subsection{Origin of the Optically-thick Absorbing Clouds}

From the observed timescales of the X-ray spectral variation and the ionization condition, MEI2012 estimated 
sizes and locations of the  partial absorbing clouds,   
and proposed that the partial covering clouds in the VPC model are the broad line region (BLR) clouds. 
We revisit this model assuming the BLR parameters estimated from optical observations.
We assume the following typical BLR parameters; size of BLR, $R\approx 2\times10^{16}$ cm, BLR cloud size, $l \approx 10^{13}$cm, 
and a BLR cloud velocity, $v\approx 5 \times 10^8$ cm s$^{-1}$ (Petterson 2006).

First, the absorbing clouds should be  optically thick at the iron K-edge, such that $N_H\approx 2 \times 10^{24}$ cm$^{-2}$,
and $\xi \lesssim 100$ (Table \ref{t2}). Ionization parameter of a BLR cloud may be estimated as follows;
\begin{equation}
\xi = \frac{L}{nR^2} \approx 0.1 \frac{(L/10^{43} {\rm \;erg\: s^{-1}}) (l/10^{13} \;{\rm cm})}{(N_H/2 \times 10^{24} \; {\rm cm^{-2}})\; (R/2\times10^{16}\;{\rm cm})^2},
\end{equation}
thus  the assumption of the optical-thickness is valid.

Next, we estimate the X-ray emission region size and the variation timescale.  We do not see either full-covering nor the no-covering, and
a typical covering fraction is $\sim0.5$ (Figure \ref{f4}). Thus, if we take the X-ray emission region size $x$, 
$(l/x)^2\approx 0.5$, namely $x \approx 1.4 \;  l = 1.4 \times 10^{13} \: {\rm cm}$.
If normalized by the Schwarzschild radius, $R_S  = 1.5 \times 10^{12}  (M/5 \times 10^6 M_\odot)$ cm, 
$x \approx  10 \; (M_\odot/5 \times 10^6 M_\odot) R_S$.  Presumably, this is reasonable as an X-ray emitting corona surrounding  the central black hole.
Also, typical flux variation timescale due to passage of a BLR cloud 
 may be estimated as $x/v \approx 3 \times 10^4$ sec, which agrees with observations (Figure \ref{f4}).

We notice that, in MCG-6-30-15 and the 20 Seyfert galaxies analyzed in this paper, such moderate 6.4 keV emission lines with typical equivalent-widths of $\sim50$ eV are observed, that are consistent with the fluorescence from the  distant cold reflectors with $\Omega/2\pi \sim 0.3$ which we assumed.  If the central X-ray sources are almost fully surrounded by 
thick and cold absorbers in these AGNs, 
much stronger fluorescent lines would have been observed  (e.g., Reynolds et al.\ 2009).
Thus, the  weakness of the observed fluorescent  emission line  suggests that 
the partial absorbing materials are directional and/or localized,  occupying a much smaller solid angle than $4 \pi$. 
It has been pointed out by theoretical simulations that accretion disks in AGNs 
tend to have disk winds or outflows, which imprints 
a variety of spectroscopic signatures including  absorption lines and edges  in the
X-ray/UV spectra (e.g., Proga and Kallman 2004, Sim et al.\ 2010, Nomura et al.\  2013). 
So, the partial covering clouds may be a part of such  fragmented and directional disk winds or outflows.
Furthermore, Doppler motion of the outflowing clouds will smear the narrow emission  line features,
which would make the fluorescent lines less noticeable.

According to the simulation by Nomura et al.\ (2013), where condition of the Broad Absorption Line (BAL)
quasars is studied, the  broad iron K-absorption lines  due to
very fast outflow ($\gtrsim 10^4$ km s$^{-1}$) of low-ionized material ($\xi \lesssim 100$) are observed
only in a narrow range of the viewing  angle between $\theta=45.\arcdeg6$  and $54.\arcdeg0$. When the viewing angle is larger
(closer to edge-on), 
the BAL will not be produced, as the outflow gas will be too thick ($N_H \gtrsim 10^{24}$ cm$^{-2}$) and the velocity will be too low.
These conditions are, however,  rather suitable for the optically-thick  partial absorbing clouds, which are required in our VPC model.
Thus, the Seyfert galaxies observed through the optically-thick partial absorbers may be the outflowing AGNs seen close to  edge-on.
In fact,  such AGN outflow gas might be   origin of the BLR clouds (Patterson 2006).

\subsection{Comments on the Relativistic Disk-line Model}
Our VPC model explains the observed broad and unvarying iron line features from Seyfert galaxies   
in terms of fluctuation  of the partial absorbers in the line of sight. 
An alternative model which might  explain the broad and unvarying iron line features 
is the ``relativistic disk-line model'' (Fabian et al.\ 2014 and references therein),
where the fluorescent iron lines are supposed to be emitted from the innermost region of the illuminated
accretion disk,     significantly skewed and broadened   by strong relativistic effects.
In this scenario, 
suppression of the iron line variability is explained by relativistic 
``light-bending effect''  (Miniutti \& Fabian 2004),
where the disk-reflected photons are less variable than the direct photons from the 
illuminating source, while   the height of a point-like illuminating  source above the accretion disk 
 varies.

Whereas the light-bending model qualitatively explains the invariability of the broad 
iron emission line,  it does  not seem to  explain the observed spectral variations  quantitatively 
(see, e.g., Goosmann et al.\ 2006, Nied\'zwiecki and Miyakawa 2010, \.Zycki et al.\ 2010, Gallo et al.\ 2015). 
The most significant parameter to cause observable spectral changes in the light-bending model 
is the height of the point-like source above the accretion disk.
So that the light-bending model to remain compelling, it should  explain quantitatively the
observed spectral variations   (Figures  \ref{f3} and  \ref{f4}), as well as 
the RMS spectra  (Figure \ref{f6}) by only change of the source height
above  the disk. 

Presumably, a   critical point of the relativistic disk-line model is its 
extreme concentration of the disk-reflected emission within the innermost region where the relativistic
effects are the strongest.
For instance, 
in extreme cases, a prediction was made based on the disk-line model that much of the X-ray emission should
originate from within only a few gravitational radii (e.g., Fabian and Vaughan 2003, Fabian et al.\ 2009).
If this is the  case, since the AGN is surrounded by many absorbing  clouds far outside, 
when an absorber greater than the X-ray emission region size
moves  across the line-of-sight,  we shall see an abrupt {\em total}\/
eclipse, but never a  gradual  {\em partial}\/ eclipse.
However,  observational evidences are being accumulated for Seyfert galaxies  that the 
 extended X-ray sources are  partially and progressively 
obscured by intervening absorbers with comparable sizes
 (e.g., McKerna \& Yaqoob 1998, Maiolino et al.\ 2010, Sanfrutos, M.\ 2013, Kaastra et al.\ 2014, Marinucci et al.\ 2014, Beuchert et al.\ 2015).  
In an occultation event identified in  MCG-6-30-15, 
the partial covering fraction varied gradually  by 0.32 within $\sim$20 ksec 
by a BLR cloud far outside  (Marinucci et al.\ 2014).
This strongly suggests  that the X-ray source is more extended than the absorber, contrary to the
assumption of the relativistic disk-line  model.

More unambiguous constraint of the X-ray emission region
size may come from the X-ray micro-lensing observations.
Chartas et al.\ (2012) observed progressive profile changes of the fluorescent
iron emission lines from the gravitationally lensed quasar RX J1131--1231, when
the caustic passes over the inner accretion disk.  The lines are distorted by
general relativistic and Doppler effects, and the line emitting 
radius is directly measured from these  effects as  $\sim$15 times the gravitational radius.
Mosquera et al.\ (2013) directly measured the X-ray emission region size of Q2237+0305;
the half-light radius of the hard X-ray (3.5 -- 21.5 keV)
is $ R= 10^{15.5\pm0.3}$ cm, which is an order of magnitude smaller than that of the optical emission, and
marginally smaller than that of the soft X-ray emission (1.1 -- 3.5  keV).
It is particularly of interest that the {\em lower-limit}\/ of the X-ray emission region size is directly
constrained.  Taking the black hole mass estimates as  $ 0.9 \times 10^9 M_\odot$ (Morgan et al.\ 2010) or   2.4$\times 10^9 M_\odot$ (Assef et al.\ 2011),  hard X-ray emission  region size (half-light radius) is
constrained as  $24\pm^{23}_{12}$  or $9\pm^{9}_{5}$ times the gravitational radius, respectively.
These  micro-lensing results 
are consistent with the partial covering scenario
that X-ray emission region is extended as much as  a few tens of the gravitational radius 
and comparable in size with the partial absorbing clouds  in the line-of-sight.
So that the relativistic reflection scenario works, 
the innermost disk region has to be  sufficiently illuminated, and  
at least a part of the X-ray emitting corona should lie within about 10 gravitational radii
 (Fabian et al.\ 2014).  
Future micro-lensing observations are expected to constrain the lower-limit of the X-ray emission region size more tightly, 
and tell if the relativistic disk-line scenario is truly feasible or not.

\section{Conclusions}\label{coclusion}
In this paper, we have studied origin of the soft X-ray spectral variation and broad  iron line feature commonly observed
in Seyfert galaxies.
We have applied the Variable Partial Covering (VPC) model, which was originally
proposed for MCG-6-30-15 (\cite{Miyakawa2012}) and has been shown to be  successful to 1H0707--495 as well \citep{Mizumoto2014}, 
to  20 other Seyfert galaxies observed with Suzaku. 
In the VPC model, most spectral variations below $\sim$10 keV 
in  a timescale within  $\sim$day are explained by variation of the partial covering
fraction of the extended X-ray source due to fluctuations of  the optically-thick absorbing clouds in the line of sight.
Not only the broad  iron line features, but also the 2 -- 10 keV spectral variations are successfully
explained by the VPC model.  
We conclude  that observed X-ray flux/spectral variation of Seyfert galaxies is explained by variation of the
partial covering fraction in timescales below $\sim$day and the  intrinsic soft X-ray luminosity variation over
$\sim10^6$ sec, both of which are mainly responsible for  $\lesssim$10 keV, and independent hard X-ray variations above $\sim$10 keV.

\bigskip

This research has made use of public Suzaku data obtained through the Data ARchives and
Transmission System (DARTS), provided by the Institute of Space and Astronautical Science (ISAS) at
Japan Aerospace Exploration Agency (JAXA).  For data reduction, we used software provided by
the High Energy Astrophysics Science Archive Research Center (HEASARC) at NASA/Goddard
Space Flight Center.  
HS is financially supported by JSPS Grant-in-Aid for JSPS Fellows Grant Number 12J10755, and
MM is  supported by JSPS KAKENHI Grant Number 15J07567.

\begin{table*}[t]
\begin{center}
\rotatebox{90}{\begin{minipage}{240mm}
\begin{center}
{
\renewcommand\arraystretch{1.}
	\caption{Best-fit parameters for time-averaged spectra$^\ast$}
	\label{t2}
\begin{tabular}{lcccccccccc}
\hline
	\phantom{OOO}(1)		&	(2)					&	(3)						&
	(4)					&	(5)					&	(6)						&
	(7)					&	(8)					&	\multicolumn{2}{c}{(9)}		&(10)\\   
	\phantom{OO}Target		&	$N_{\rm H,{\rm ISM}}$	&	$N_{{\rm H},H}$			&
	$N_{{\rm H},L}$		&	$N_{1}$				&	$N_{2}$&$N_{{\rm H},2}$		&
	$\Gamma$&$E_{\rm Fe}$	&	$I_{\rm Fe}$			&	Re-${\chi^{2}}$\\
						&						&	log $\xi_{H}$				&
	log $\xi_{L}$			&						&							&
	log $\xi_{2}$			&&	\multicolumn{2}{c}{$EW_{\rm Fe}$}						&(d.o.f)\\
\hline
Markarian\,335  & $0.28^{+0.15}_{-0.17}$ & $< 0.01$\phantom{$^{00000000}$} & $< 0.1$\phantom{$0^{00000000}$} & \phantom{0}$7.77^{+0.51}_{-0.55}$ & \phantom{0}$7.66^{+2.54\phantom{0}}_{-2.06\phantom{0}}$ & $1.79^{+0.34}_{-0.35}$ & $2.24^{+0.04}_{-0.05}$ & $6.21^{+0.02}_{-0.03}$ & $0.64^{+0.14}_{-0.14}$ & 1.07 \\
&& -- & -- & & & $1.90^{+0.10}_{-0.26}$ &&\multicolumn{2}{c}{$40^{+\phantom{0}9}_{-\phantom{0}9}$}& (1167) \\
\hline
TonS\,180 & $< 0.01$\phantom{$^{00000000}$} & $< 0.01$\phantom{$^{00000000}$} & $< 0.1$\phantom{$0^{00000000}$} & \phantom{0}$3.53^{+0.13}_{-0.13}$ & \phantom{0}$9.86^{+4.43\phantom{0}}_{-4.11\phantom{0}}$ & $4.51^{+0.86}_{-1.85}$ & $2.35^{+0.03}_{-0.03}$ & \phantom{$^{0}$}$6.16^{(fixed)}$ & $0.08^{+0.10}_{-0.08}$ & 1.02 \\
&& -- & -- & & & $2.04^{+0.30}_{-0.68}$ &&\multicolumn{2}{c}{$< 32$\phantom{$^{000000}$}}& (\phantom{$0$}425) \\
\hline
1H\,0323+342  & $< 0.01$\phantom{$^{00000000}$} & $< 0.01$\phantom{$^{00000000}$} & $< 0.1$\phantom{$0^{00000000}$} & \phantom{0}$3.33^{+0.14}_{-0.14}$ & \phantom{0}$1.44^{+0.95\phantom{0}}_{-0.87\phantom{0}}$ & $1.81^{+2.81}_{-0.44}$ & $1.90^{+0.04}_{-0.04}$ & \phantom{$^{0}$}$6.40^{(fixed)}$ & $0.01^{+0.18}_{-0.01}$ &  0.93 \\
&& -- & -- & & & $1.40^{+0.80}_{-0.09}$ &&\multicolumn{2}{c}{$< 17$\phantom{$^{000000}$}}& (\phantom{$0$}498) \\
\hline
1H\,0419--577 (1) & $< 0.01$\phantom{$^{00000000}$} & $0.23^{+0.60}_{-0.18}$ & $< 0.1$\phantom{$0^{00000000}$} &\phantom{0} $5.15^{+0.11}_{-0.04}$ & \phantom{0}$3.00^{+0.73\phantom{0}}_{-0.67\phantom{0}}$ & $2.15^{+0.55}_{-0.17}$ & $1.86^{+0.02}_{-0.01}$ & $5.79^{+0.06}_{-0.05}$ & $0.21^{+0.13}_{-0.13}$ & 1.06 \\
&& $3.47^{+0.44}_{-0.24}$ & -- & & & $1.65^{+0.44}_{-0.27}$ &&\multicolumn{2}{c}{$\phantom{0}10^{+\phantom{0}5}_{-\phantom{0}6}$}& (1280) \\
\hline
1H\,0419--577 (2) & $< 0.01$\phantom{$^{00000000}$} & $< 0.01$\phantom{$^{00000000}$} & $< 0.1$\phantom{$0^{00000000}$} & \phantom{0}$4.01^{+0.11}_{-0.12}$ & \phantom{0}$1.17^{+0.68\phantom{0}}_{-0.54\phantom{0}}$ & $1.92^{+1.68}_{-0.52}$ & $1.83^{+0.03}_{-0.03}$ & $5.79^{+0.03}_{-0.03}$ & $0.41^{+0.18}_{-0.18}$ & 1.07 \\
&& -- & -- & & & $1.48^{+0.80}_{-0.25}$ &&\multicolumn{2}{c}{$\phantom{0}24^{+\phantom{0}9}_{-\phantom{0}9}$}& (\phantom{$0$}706) \\
\hline
Arakelian\,120  & $< 0.01$\phantom{$^{00000000}$} & $< 0.01$\phantom{$^{00000000}$} & $< 0.1$\phantom{$0^{00000000}$} & $11.29^{+0.36}_{-0.23}$ & \phantom{0}$7.90^{+1.76\phantom{0}}_{-1.20\phantom{0}}$ & $2.74^{+0.18}_{-1.08}$ & $2.02^{+0.03}_{-0.02}$ & $6.20^{+0.02}_{-0.01}$ & $1.86^{+0.29}_{-0.27}$ & 0.90 \\
&& -- & -- & & & $1.13^{+0.24}_{-1.13}$ &&\multicolumn{2}{c}{$\phantom{0}56^{+11}_{-\phantom{0}8}$}& (1068) \\
\hline
NGC\,3516  & $1.32^{+0.16}_{-0.09}$ & $2.64^{+8.65}_{-1.53}$ & $< 0.1$\phantom{$0^{00000000}$} & \phantom{0}$3.17^{+0.20}_{-0.21}$ & \phantom{0}$2.17^{+0.61\phantom{0}}_{-0.58\phantom{0}}$ & $1.55^{+0.27}_{-0.19}$ & $1.69^{+0.04}_{-0.05}$ & $6.34^{+0.00}_{-0.00}$ & $3.66^{+0.17}_{-0.17}$ & 0.98 \\
&& $3.48^{+0.48}_{-0.08}$ & -- & & & $1.40^{+0.46}_{-0.10}$ &&\multicolumn{2}{c}{$224^{+13}_{-13}$}& (1153) \\
\hline
NGC\,3783 (1) & $0.93^{+0.22}_{-0.13}$ & $0.13^{+0.19}_{-0.05}$ & $< 0.1$\phantom{$0^{00000000}$} & $11.10^{+0.99}_{-0.53}$ & \phantom{0}$3.80^{+2.20\phantom{0}}_{-0.87\phantom{0}}$ & $1.31^{+0.32}_{-0.36}$ & $1.70^{+0.06}_{-0.03}$ & $6.33^{+0.01}_{-0.01}$ & $6.34^{+0.36}_{-0.37}$ & 0.98 \\
&& $2.94^{+0.20}_{-0.09}$ & -- & & & $1.16^{+0.96}_{-0.41}$ &&\multicolumn{2}{c}{$117^{+\phantom{0}7}_{-\phantom{0}9}$}& (1561) \\
\hline
NGC\,3783 (2) & $0.27^{+0.44}_{-0.27}$ & $0.33^{+0.23}_{-0.15}$ & $2.51^{+1.02}_{-0.97}$ & $16.72^{+0.88}_{-0.83}$ & \phantom{0}$5.85^{+1.65\phantom{0}}_{-1.30\phantom{0}}$ & $1.74^{+0.85}_{-0.27}$ & $1.74^{+0.03}_{-0.03}$ & $6.32^{+0.01}_{-0.00}$ & $6.10^{+0.41}_{-0.34}$ & 1.08 \\
&& $3.04^{+0.09}_{-0.13}$ & $1.91^{+0.13}_{-0.10}$ & & & $0.28^{+0.98}_{-0.28}$ &&\multicolumn{2}{c}{$\phantom{0}86^{+\phantom{0}7}_{-\phantom{0}6}$}& (1789) \\
\hline
NGC\,4051  & $< 0.01$\phantom{$^{00000000}$} & $0.07^{+0.11}_{-0.05}$ & $< 0.1$\phantom{$0^{00000000}$} & \phantom{0}$8.69^{+0.10}_{-0.14}$ & \phantom{0}$6.93^{+0.80\phantom{0}}_{-0.90\phantom{0}}$ & $2.34^{+0.71}_{-0.37}$ & $2.00^{+0.01}_{-0.01}$ & $6.39^{+0.01}_{-0.01}$ & $1.50^{+0.13}_{-0.13}$  & 1.11 \\
&& $3.20^{+0.18}_{-0.21}$ & -- & & & $0.45^{+0.10}_{-0.05}$ &&\multicolumn{2}{c}{$\phantom{0}63^{+\phantom{0}6}_{-\phantom{0}6}$}& (1666) \\
\hline
Markarian\,205  & $0.33^{+0.31}_{-0.33}$ & $< 0.01$\phantom{$^{00000000}$} & $< 0.1$\phantom{$0^{00000000}$} & \phantom{0}$3.40^{+0.46}_{-0.46}$ & \phantom{0}$6.55^{+3.21\phantom{0}}_{-2.44\phantom{0}}$ & $5.11^{+1.54}_{-0.41}$ & $1.98^{+0.09}_{-0.10}$ & $5.97^{+0.03}_{-0.03}$ & $0.71^{+0.18}_{-0.18}$ & 1.05 \\
&& -- & -- & & & $2.87^{+0.09}_{-0.56}$ &&\multicolumn{2}{c}{$\phantom{0}65^{+15}_{-23}$}& (\phantom{$0$}425) \\
\hline
Markarian\,766  & $0.18^{+0.45}_{-0.18}$ & $< 0.01$\phantom{$^{00000000}$} & $2.66^{+2.36}_{-2.66}$ & \phantom{0}$6.70^{+0.85}_{-0.35}$ & \phantom{0}$5.08^{+2.55\phantom{0}}_{-2.09\phantom{0}}$ & $2.15^{+2.32}_{-0.66}$ & $2.16^{+0.07}_{-0.09}$ & $6.39^{+0.02}_{-0.03}$ & $1.06^{+0.39}_{-0.33}$ & 1.02 \\
&& -- & $2.16^{+0.31}_{-0.57}$ & & & $1.55^{+0.74}_{-1.55}$ &&\multicolumn{2}{c}{$\phantom{0}62^{+61}_{-27}$}& (\phantom{$0$}663) \\
\hline
  \multicolumn{10}{@{}l@{}}{\hbox to 0pt{\parbox{220mm}{\footnotesize 
  \noindent 
  }\hss}}
\end{tabular}
}
\end{center}
\end{minipage}}
\end{center}
\end{table*}
\addtocounter{table}{-1}
\begin{table*}[t]
\begin{center}
\rotatebox{90}{\begin{minipage}{240mm}
\begin{center}
{
\renewcommand\arraystretch{1.}
	\caption{Best-fit parameters for time-averaged spectra (continued)}
	\label{t2}
\begin{tabular}{lcccccccccc}
\hline
	\phantom{OOO}(1)		&	(2)					&	(3)						&
	(4)					&	(5)					&	(6)						&
	(7)					&	(8)					&	\multicolumn{2}{c}{(9)}		&(10)\\   
	\phantom{OO}Target		&	$N_{\rm H,{\rm ISM}}$	&	$N_{{\rm H},H}$			&
	$N_{{\rm H},L}$		&	$N_{1}$				&	$N_{2}$&$N_{{\rm H},2}$		&
	$\Gamma$&$E_{\rm Fe}$	&	$I_{\rm Fe}$			&	Re-${\chi^{2}}$\\
						&						&	log $\xi_{H}$				&
	log $\xi_{L}$			&						&							&
	log $\xi_{2}$			&&	\multicolumn{2}{c}{$EW_{\rm Fe}$}						&(d.o.f)\\
\hline
NGC\,4593  & $0.43^{+0.17}_{-0.14}$ & $< 0.01$\phantom{$^{00000000}$} & $< 0.1$\phantom{$0^{00000000}$} & \phantom{0}$2.44^{+0.25}_{-0.22}$ & \phantom{0}$1.69^{+0.90\phantom{0}}_{-0.69\phantom{0}}$ & $1.80^{+1.07}_{-0.36}$ & $1.73^{+0.06}_{-0.06}$ & $6.35^{+0.01}_{-0.01}$ & $2.56^{+0.18}_{-0.19}$ & 1.04 \\
&& -- & -- & & & $1.40^{+0.50}_{-0.34}$ &&\multicolumn{2}{c}{$222^{+18}_{-22}$}& (\phantom{$0$}600) \\
\hline
IC\,4329A  & $0.61^{+0.09}_{-0.04}$ & $< 0.01$\phantom{$^{00000000}$} & $< 0.1$\phantom{$0^{00000000}$} & $32.84^{+0.47}_{-0.49}$ & $12.35^{+1.90\phantom{0}}_{-1.13\phantom{0}}$ & $1.62^{+0.27}_{-0.07}$ & $1.85^{+0.01}_{-0.02}$ & $6.29^{+0.01}_{-0.01}$ & $5.62^{+0.45}_{-0.42}$ & 1.01 \\
&& -- & -- & & & $1.49^{+0.23}_{-0.52}$ &&\multicolumn{2}{c}{$\phantom{0}47^{+\phantom{0}3}_{-\phantom{0}4}$}& (1877) \\
\hline
NGC\,5548  & $0.41^{+0.05}_{-0.09}$ & $0.13^{+0.07}_{-0.05}$ & $< 0.1$\phantom{$0^{00000000}$} & \phantom{0}$4.25^{+0.25}_{-0.09}$ & \phantom{0}$1.43^{+0.27\phantom{0}}_{-0.27\phantom{0}}$ & $1.63^{+0.30}_{-0.32}$ & $1.67^{+0.03}_{-0.03}$ & $6.29^{+0.01}_{-0.01}$ & $1.78^{+0.15}_{-0.14}$ & 1.05 \\
&& $2.97^{+0.17}_{-0.08}$ & -- & & & $0.36^{+0.04}_{-0.05}$ &&\multicolumn{2}{c}{$\phantom{0}87^{+\phantom{0}3}_{-13}$}& (1435) \\
\hline
4C\,74.26   & $0.59^{+0.18}_{-0.18}$ & $< 0.01$\phantom{$^{00000000}$} & $< 0.1$\phantom{$0^{00000000}$} & $11.75^{+0.95}_{-0.85}$ & \phantom{0}$5.03^{+2.70\phantom{0}}_{-2.11\phantom{0}}$ & $2.72^{+0.61}_{-0.39}$ & $1.98^{+0.05}_{-0.05}$ & $5.81^{+0.03}_{-0.03}$ & $1.32^{+0.31}_{-0.30}$ & 1.05 \\
&& -- & -- & & & $2.66^{+0.12}_{-0.15}$ &&\multicolumn{2}{c}{$\phantom{0}34^{+\phantom{0}7}_{-\phantom{0}8}$}& (1007) \\
\hline
Arakelian\,564  & $0.22^{+0.22}_{-0.18}$ & $0.44^{+3.87}_{-0.41}$ & $< 0.1$\phantom{$0^{00000000}$} & $16.71^{+1.12}_{-1.63}$ & $23.51^{+10.38}_{-8.03\phantom{0}}$ & $2.78^{+1.42}_{-0.56}$ & $2.60^{+0.07}_{-0.06}$ & $6.43^{+0.06}_{-0.20}$ & $0.28^{+0.48}_{-0.21}$ & 0.98 \\
&& $3.38^{+0.68}_{-1.16}$ & -- & & & $1.99^{+0.34}_{-0.48}$ &&\multicolumn{2}{c}{$\phantom{0}18^{+13}_{-15}$}& (\phantom{$0$}770) \\
\hline
SWIFT\,J2127.4  & $1.57^{+0.24}_{-0.25}$ & $< 0.01$\phantom{$^{00000000}$} & $< 0.1$\phantom{$0^{00000000}$} & $17.13^{+1.65}_{-1.70}$ & $12.13^{+6.06\phantom{0}}_{-4.48\phantom{0}}$ & $1.66^{+0.37}_{-0.42}$ & $2.17^{+0.07}_{-0.07}$ & $6.28^{+0.04}_{-0.04}$ & $0.98^{+0.44}_{-0.44}$ & 0.94 \\
&& -- & -- & & & $1.95^{+0.20}_{-0.74}$ &&\multicolumn{2}{c}{$\phantom{0}26^{+10}_{-12}$}& (\phantom{$0$}645) \\
\hline
NGC\,7469  & $0.22^{+0.16}_{-0.22}$ & $0.07^{+0.37}_{-0.07}$ & $< 0.1$\phantom{$0^{00000000}$} & \phantom{0}$5.94^{+0.34}_{-0.35}$ & \phantom{0}$3.08^{+1.35\phantom{0}}_{-1.21\phantom{0}}$ & $1.64^{+1.06}_{-0.28}$ & $1.83^{+0.03}_{-0.05}$ & $6.27^{+0.01}_{-0.01}$ & $2.47^{+0.25}_{-0.25}$ & 0.93 \\
&& $2.99^{+0.41}_{-0.32}$ & -- & & & $1.42^{+0.58}_{-0.44}$ &&\multicolumn{2}{c}{ $107^{+11}_{-13}$}& (\phantom{$0$}875) \\
\hline
NGC\,2992  & $1.84^{+0.14}_{-0.24}$ & $< 0.01$\phantom{$^{00000000}$} & $< 0.1$\phantom{$0^{00000000}$} & \phantom{0}$3.34^{+0.24}_{-0.32}$ & \phantom{0}$1.80^{+0.66\phantom{0}}_{-0.69\phantom{0}}$ & $1.30^{+0.48}_{-0.22}$ & $1.80^{+0.04}_{-0.07}$ & $6.34^{+0.01}_{-0.00}$ & $3.11^{+0.17}_{-0.17}$& 1.05 \\
&& -- & -- & & & $1.39^{+0.37}_{-0.41}$ &&\multicolumn{2}{c}{$218^{+14}_{-15}$ }& (\phantom{$0$}860) \\
\hline
MCG-5-23-16  & $2.40^{+0.06}_{-0.08}$ & $< 0.01$\phantom{$^{00000000}$} & $< 0.1$\phantom{$0^{00000000}$} & $29.65^{+0.68}_{-0.94}$ & $11.08^{+1.46\phantom{0}}_{-2.14\phantom{0}}$ & $1.50^{+0.14}_{-0.12}$ & $1.85^{+0.01}_{-0.02}$ & $6.33^{+0.01}_{-0.01}$ & $7.35^{+0.42}_{-0.42}$ & 1.08 \\
&& -- & -- & & & $1.39^{+0.12}_{-0.06}$ &&\multicolumn{2}{c}{$\phantom{0}68^{+\phantom{0}4}_{-\phantom{0}4}$}& (1829) \\
\hline
IRAS\,18325  & $2.48^{+0.21}_{-0.22}$ & $< 0.01$\phantom{$^{00000000}$} & $< 0.1$\phantom{$0^{00000000}$} & $19.83^{+1.32}_{-1.80}$ & $10.35^{+4.37\phantom{0}}_{-3.91\phantom{0}}$ & $1.57^{+0.30}_{-0.53}$ & $2.41^{+0.06}_{-0.06}$ & $6.28^{+0.03}_{-0.03}$ & $0.80^{+0.30}_{-0.30}$ & 1.02 \\
&& -- & -- & & & $1.90^{+0.16}_{-1.30}$ &&\multicolumn{2}{c}{$\phantom{0}29^{+12}_{-11}$}& (\phantom{$0$}817) \\
\hline
  \multicolumn{10}{@{}l@{}}{\hbox to 0pt{\parbox{220mm}{\footnotesize 
  \noindent 
  \footnotemark[$*$] Errors correspond to 90 \% confidence limits.
Meaning of the tables columns are the following: (1) Target name. (2) Hydrogen column-density of the interstellar matter ($10^{22}$ cm$^{-2}$). 
 (3) Hydrogen column density ($10^{23}$ cm$^{-2}$) and  logarithm of the ionization parameter for $W_H$.  (4) Hydrogen column density ($10^{22}$ cm$^{-2}$) and  logarithm of the ionization parameter for $W_L$.  (5) Normalization of the direct component
(10$^{-3}$ photons s$^{-1}$ cm$^{-1}$ keV$^{-1}$ at 1 keV). (6)
Normalization of the absorbed component
(10$^{-3}$ photons s$^{-1}$ cm$^{-1}$ keV$^{-1}$ at 1 keV). (7) 
Hydrogen column-density  ($10^{24}$ cm$^{-2}$) and
logarithm of the ionization parameter
for $W_2$. (8) Photon-index of the power-law component.  (9) Iron line energy
(keV), intensity (10$^{-5}$ photons s$^{-1}$ cm$^{-2}$) and equivalent-width (eV). (10) Reduced $\chi^2$ and the degree of freedom.
  }\hss}}
\end{tabular}
}
\end{center}
\end{minipage}}
\end{center}
\end{table*}
\begin{center}
\begin{table*}[t]
\begin{center}
{\tabcolsep = 1.5truemm
\renewcommand\arraystretch{1.}
	\caption{Best-fit parameters for the intensity-sliced spectra$^\ast$}
	\label{t3}
\begin{tabular}{lccccccccc}
\hline
        \phantom{OOOO}(1) & (2) & (3) & (4)  && (5) & (6) & (7) & (8) & (9)\\
        \multirow{4}{*}{\phantom{OOO}Target} & \multirow{4}{*}{$N_{\rm H,{\rm ISM}}$} & & \multirow{4}{*}{$N$} & & $\alpha_{1}$ &   &  & \multirow{4}{*}{$\Gamma$} &\\
       &  & $N_{{\rm H},H}$ & && $\alpha_{2}$ & $<N_{{\rm H},L}>$ & $N_{{\rm H},2}$ & & Re-${\chi^{2}}$ \\
       &  &  log $\xi_{H}$ && & $\alpha_{3}$ & log $\xi_{L}$&log $\xi_{2}$&&(d.o.f)\\
       &  &  && & $\alpha_{4}$ &&&&\\
\hline
\multirow{4}{*}{Markarian\,335}  & \multirow{4}{*}{$0.20^{+0.08}_{-0.08}$} &  & \multirow{4}{*}{$1.48$} && $0.61^{+0.11}_{-0.09}$  & \multirow{2}{*}{$< 0.1$\phantom{$0^{00000000}$}} &  & \multirow{4}{*}{$2.19^{+0.02}_{-0.01}$} &  \\
&& $< 0.01$\phantom{$^{00000000}$}&  && $0.53^{+0.13}_{-0.11}$ & & $1.79$\phantom{$^{+0.00}$} && 1.03 \\
       &  & -- & &&  $0.46^{+0.16}_{-0.12}$ & -- &$1.65^{+0.08}_{-0.10}$ & & (1709)\\
       &  &  & && $0.35^{+0.19}_{-0.14}$ & & & &\\
\hline
\multirow{4}{*}{TonS\,180} & \multirow{4}{*}{$< 0.01$\phantom{$^{00000000}$}} &  & \multirow{4}{*}{$1.33$} && $0.78^{+0.08}_{-0.08}$ && & \multirow{4}{*}{$2.33^{+0.03}_{-0.03}$} & \\
&& $< 0.01$\phantom{$^{00000000}$} &  && $0.74^{+0.09}_{-0.09}$ &  $< 0.1$\phantom{$0^{00000000}$}  & $4.51$\phantom{$^{+0.00}$} &&0.90 \\
       &  & -- & && $0.73^{+0.10}_{-0.10}$ & -- & $1.97^{+0.14}_{-0.70}$& &(\phantom{0}476)\\
       &  &  & && $0.67^{+0.12}_{-0.12}$ & & & &\\
\hline
\multirow{4}{*}{1H\,0323+342}  & \multirow{4}{*}{$< 0.01$\phantom{$^{00000000}$}} &  & \multirow{4}{*}{$0.51$} && $0.46^{+0.09}_{-0.10}$ & & & \multirow{4}{*}{$1.89^{+0.02}_{-0.01}$} &  \\
&& $< 0.01$\phantom{$^{00000000}$}  &  && $0.37^{+0.11}_{-0.10}$ & $< 0.1$\phantom{$0^{00000000}$}  & $1.81$\phantom{$^{+0.00}$}  && 0.97 \\
       &  & -- & && $0.31^{+0.12}_{-0.12}$ & -- & $0.36^{+0.03}_{-0.04}$ & & (1357)\\
       &  &  & && $0.26^{+0.13}_{-0.13}$ & & & &\\
\hline
\multirow{4}{*}{1H\,0419--577 (1)} & \multirow{4}{*}{$< 0.01$\phantom{$^{00000000}$}} &  & \multirow{4}{*}{$0.80$} && $0.44^{+0.04}_{-0.03}$  &  & & \multirow{4}{*}{$1.86^{+0.01}_{-0.01}$} &  \\
&&$0.23$\phantom{$^{+0.00}$} &  && $0.38^{+0.05}_{-0.06}$ & $< 0.1$\phantom{$0^{00000000}$} & $2.15$\phantom{$^{+0.00}$} &&1.01  \\
       &  & $3.51^{+0.37}_{-0.16}$  & && $0.32^{+0.05}_{-0.08}$ &-- &$1.65^{+0.07}_{-0.08}$ & &(1877)\\
       &  &  & && $0.27^{+0.06}_{-0.08}$ & & & &\\
\hline
\multirow{4}{*}{1H\,0419--577 (2)} & \multirow{4}{*}{$< 0.01$\phantom{$^{00000000}$}} &  & \multirow{4}{*}{$0.53$} && $0.31^{+0.09}_{-0.13}$ &   & & \multirow{4}{*}{$1.83^{+0.01}_{-0.01}$} &  \\
&& $< 0.01$\phantom{$^{00000000}$} &  && $0.26^{+0.08}_{-0.11}$ & $< 0.1$\phantom{$0^{00000000}$}  & $1.92$\phantom{$^{+0.00}$} && 1.07 \\
       &  & -- && & $0.21^{+0.09}_{-0.12}$ & -- & $1.65^{+0.16}_{-0.74}$& &(\phantom{0}867)\\
       &  &  & && $0.19^{+0.12}_{-0.13}$ & & & &\\
\hline
\multirow{4}{*}{Arakelian\,120}  & \multirow{4}{*}{$< 0.01$\phantom{$^{00000000}$}} && \multirow{4}{*}{$1.90$} && $0.48^{+0.08}_{-0.08}$  &  && \multirow{4}{*}{$2.01^{+0.01}_{-0.01}$} &  \\
&&  $< 0.01$\phantom{$^{00000000}$} & &&  $0.44^{+0.09}_{-0.09}$ & $< 0.1$\phantom{$0^{00000000}$} & $2.74$\phantom{$^{+0.00}$}  && 1.00 \\
       &  & -- && & $0.38^{+0.10}_{-0.10}$ & -- &$0.99^{+0.27}_{-0.58}$ & &(1383)\\
       &  &  & && $0.33^{+0.11}_{-0.11}$ & & & &\\
\hline
\multirow{4}{*}{NGC\,3783 (1)} & \multirow{4}{*}{$1.09^{+0.07}_{-0.07}$} &  & \multirow{4}{*}{$1.69$} && $0.39^{+0.09}_{-0.09}$ &  & & \multirow{4}{*}{$1.73^{+0.01}_{-0.01}$} &  \\
&& $0.13$\phantom{$^{+0.00}$}&  && $0.33^{+0.10}_{-0.05}$ & $< 0.1$\phantom{$0^{00000000}$} & $1.31$\phantom{$^{+0.00}$} && 1.05\\
       &  & $2.94^{+0.10}_{-0.08}$  & && $0.28^{+0.11}_{-0.11}$ & -- &$1.35^{+0.19}_{-0.06}$ & &(2447) \\
       &  &  & && $0.19^{+0.13}_{-0.13}$ & & & &\\
\hline
\multirow{4}{*}{NGC\,3783 (2)} & \multirow{4}{*}{$1.30^{+0.06}_{-0.14}$} &  & \multirow{4}{*}{$2.10$} && $0.36^{+0.07}_{-0.09}$ &  &  & \multirow{4}{*}{$1.72^{+0.01}_{-0.01}$} &  \\
&& $0.33$\phantom{$^{+0.00}$} & && $0.26^{+0.07}_{-0.07}$ & $3.39$\phantom{$^{+0.00}$}& $1.74$\phantom{$^{+0.00}$} && 1.04 \\
       &  & $3.07^{+0.07}_{-0.12}$ && & $0.20^{+0.08}_{-0.11}$ &\phantom{$^{00}$}$2.28 <$\phantom{$^{00000}$}  & $0.24^{+0.04}_{-0.04}$& & (4417)\\
       &  &  && & $0.14^{+0.08}_{-0.09}$  & &  & &\\
\hline
  \multicolumn{10}{@{}l@{}}{\hbox to 0pt{\parbox{170mm}{\footnotesize 
  \noindent 
  \footnotemark[$*$] Errors correspond to 90 \% confidence limits.
Meaning of the tables columns are the following: (1) Target name. (2) Hydrogen column-density of the interstellar matter ($10^{22}$ cm$^{-2}$). 
 (3) Hydrogen column density ($10^{23}$ cm$^{-2}$) and  logarithm of the ionization parameter for $W_H$.  (4) Normalization of the power-law component
(10$^{-3}$ photons s$^{-1}$ cm$^{-1}$ keV$^{-1}$ at 1 keV). 
(5) Four partial covering fractions corresponding to the four intensity-sliced
spectra.
 (6) Common  hydrogen column density ($10^{22}$ cm$^{-2}$) of the
low-ionized absorber  and  
logarithm of the ionization parameter for $W_L$. 
 (7) 
Hydrogen column-density  ($10^{24}$ cm$^{-2}$) and
logarithm of the ionization parameter
for $W_2$. (8) Photon-index of the power-law component.   (9) Reduced $\chi^2$ and the degree of freedom.
  }\hss}}
\end{tabular}
}
\end{center}
\end{table*}
\end{center}
\addtocounter{table}{-1}
\begin{table*}[t]
\begin{center}
{\tabcolsep = 1.5truemm
\renewcommand\arraystretch{1.}
	\caption{Best-fit parameters for the intensity-sliced spectra (continued)}
	\label{t3}
\begin{tabular}{lccccccccc}
\hline
        \phantom{OOOO}(1) & (2) & (3) & (4)  && (5) & (6) & (7) & (8) & (9)\\
        \multirow{4}{*}{\phantom{OOO}Target} & \multirow{4}{*}{$N_{\rm H,{\rm ISM}}$} & & \multirow{4}{*}{$N$} & & $\alpha_{1}$ &   &  & \multirow{4}{*}{$\Gamma$} &\\
       &  & $N_{{\rm H},H}$ & && $\alpha_{2}$ & $<N_{{\rm H},L}>$ & $N_{{\rm H},2}$ & & Re-${\chi^{2}}$ \\
       &  &  log $\xi_{H}$ && & $\alpha_{3}$ & log $\xi_{L}$&log $\xi_{2}$&&(d.o.f)\\
       &  &  && & $\alpha_{4}$ &&&&\\
\hline
\multirow{4}{*}{NGC\,3516}  & \multirow{4}{*}{$1.45^{+0.06}_{-0.08}$} &  & \multirow{4}{*}{$0.58$} && $0.64^{+0.02}_{-0.08}$ & & & \multirow{4}{*}{$1.70^{+0.01}_{-0.01}$} &  \\
&& $2.64$\phantom{$^{+0.00}$} &  && $0.42^{+0.06}_{-0.05}$ &  $< 0.1$\phantom{$0^{00000000}$} & $1.55$\phantom{$^{+0.00}$} && 1.12 \\
       &  & $3.48^{+0.21}_{-0.13}$ & && $0.32^{+0.07}_{-0.07}$ & -- &$1.65^{+0.01}_{-0.03}$ & & (1448)\\
       &  &  & && $0.19^{+0.07}_{-0.07}$ & & & &\\
\hline
\multirow{4}{*}{NGC\,4051} & \multirow{4}{*}{$< 0.01$\phantom{$^{00000000}$}} &  & \multirow{4}{*}{$1.55$} && $0.66^{+0.03}_{-0.01}$  &  & & \multirow{4}{*}{$2.00^{+0.01}_{-0.01}$} &  \\
&& $0.07$\phantom{$^{+0.00}$} & && $0.47^{+0.09}_{-0.04}$ & $< 0.1$\phantom{$0^{00000000}$} & $2.34$\phantom{$^{+0.00}$}  && 1.10 \\
       &  & $3.20^{+0.18}_{-0.08}$ && & $0.32^{+0.15}_{-0.10}$ & -- &$0.43^{+0.02}_{-0.02}$  & & (4870)\\
       &  &  & && $0.08^{+0.28}_{-0.07}$ & & & &\\
\hline
\multirow{4}{*}{Markarian\,205}  & \multirow{4}{*}{$0.26^{+0.17}_{-0.16}$} &  & \multirow{4}{*}{$0.92$} && $0.67^{+0.16}_{-0.15}$ &  &  & \multirow{4}{*}{$1.97^{+0.02}_{-0.02}$} &  \\
&&$< 0.01$\phantom{$^{00000000}$} &  && $0.64^{+0.17}_{-0.16}$ & $< 0.1$\phantom{$0^{00000000}$} & $5.11$\phantom{$^{+0.00}$}  && 1.21 \\
       &  & --  && & $0.62^{+0.18}_{-0.16}$ &-- & $2.86^{+0.09}_{-0.58}$ & &(\phantom{0}479)\\
       &  &  && & $0.62^{+0.19}_{-0.17}$ & & & &\\
\hline
\multirow{4}{*}{Markarian\,766}  & \multirow{4}{*}{$0.10^{+0.19}_{-0.09}$} &  & &\ldelim\{{3}{0.3pt}[]& $0.62^{+0.15}_{-0.14}$ &  & & \multirow{4}{*}{$2.15^{+0.02}_{-0.01}$} & \\
&&$< 0.01$\phantom{$^{00000000}$} & 1.12 & &$0.44^{+0.23}_{-0.19}$ & $7.59$\phantom{$^{+0.00}$} & $2.15$\phantom{$^{+0.00}$}   && 1.08  \\
       &  & -- &&  & $0.31^{+0.27}_{-0.21}$ & $2.16^{+0.12}_{-0.03}$ &$1.65^{+0.18}_{-0.26}$ & &(\phantom{0}812)\\
       &  &  & $1.44$ &\hspace{1truemm}\ldelim\{{1}{1pt}[]& $0.18^{+0.39}_{-0.08}$ & & & &\\ 
\hline
\multirow{4}{*}{NGC\,4593}  & \multirow{4}{*}{$0.46^{+0.10}_{-0.14}$} &  & \multirow{4}{*}{$0.43$} && $0.57^{+0.07}_{-0.15}$ &  &  & \multirow{4}{*}{$1.72^{+0.02}_{-0.01}$ }& \\
&& $< 0.01$\phantom{$^{00000000}$} & && $0.46^{+0.08}_{-0.18}$ &$< 0.1$\phantom{$0^{00000000}$} & $1.80$\phantom{$^{+0.00}$} && 0.97  \\
       &  & -- && & $0.37^{+0.09}_{-0.20}$ & -- &$1.65^{+0.10}_{-0.23}$ & & (\phantom{0}699)\\
       &  &  & && $0.27^{+0.06}_{-0.08}$ & & & &\\
\hline
\multirow{4}{*}{IC\,4329A}  & \multirow{4}{*}{$0.61^{+0.02}_{-0.03}$} &  & \multirow{4}{*}{$4.40$} && $0.45^{+0.03}_{-0.02}$ &  &  & \multirow{4}{*}{$1.84^{+0.00}_{-0.00}$} &  \\
&& $< 0.01$\phantom{$^{00000000}$} & && $0.23^{+0.03}_{-0.03}$ & $< 0.1$\phantom{$0^{00000000}$} & $1.62$\phantom{$^{+0.00}$} & &1.04\\
       &  & -- && & $0.17^{+0.03}_{-0.03}$ & -- & $1.65^{+0.02}_{-0.03}$ & & (6262)\\
       &  &  && & $0.09^{+0.03}_{-0.04}$ & & & &\\
\hline
\multirow{4}{*}{NGC\,5548}  & \multirow{4}{*}{$0.62^{+0.05}_{-0.05}$} &  & \multirow{2}{*}{$0.59$} &\ldelim\{{2}{0.3pt}[]& $0.55^{+0.04}_{-0.03}$ & & & \multirow{4}{*}{$1.70^{+0.01}_{-0.01}$} &  \\
&& $0.13$\phantom{$^{+0.00}$} & && $0.24^{+0.06}_{-0.05}$ & $< 0.1$\phantom{$0^{00000000}$}  &$1.63$\phantom{$^{+0.00}$} &&1.00 \\
       &  & $2.94^{+0.16}_{-0.05}$ & \multirow{2}{*}{1.00} &\ldelim\{{2}{0.3pt}[]& $0.37^{+0.05}_{-0.04}$ & -- &  $0.36^{+0.01}_{-0.01}$ & & (3024) \\
       &  &  & && $0.14^{+0.06}_{-0.06}$ & & & &\\
\hline
  \multicolumn{10}{@{}l@{}}{\hbox to 0pt{\parbox{180mm}{\footnotesize 
  \noindent 
  }\hss}}
\end{tabular}
}
\end{center}
\end{table*}
\addtocounter{table}{-1}
\begin{table*}[t]
\begin{center}
{\tabcolsep = 1.5truemm
\renewcommand\arraystretch{1.}
	\caption{Best-fit parameters for the intensity-sliced spectra (continued)}
	\label{t3}
\begin{tabular}{lccccccccc}
\hline
        \phantom{OOOO}(1) & (2) & (3) & (4)  && (5) & (6) & (7) & (8) & (9)\\
        \multirow{4}{*}{\phantom{OOO}Target} & \multirow{4}{*}{$N_{\rm H,{\rm ISM}}$} & & \multirow{4}{*}{$N$} & & $\alpha_{1}$ &   &  & \multirow{4}{*}{$\Gamma$} &\\
       &  & $N_{{\rm H},H}$ & && $\alpha_{2}$ & $<N_{{\rm H},L}>$ & $N_{{\rm H},2}$ & & Re-${\chi^{2}}$ \\
       &  &  log $\xi_{H}$ && & $\alpha_{3}$ & log $\xi_{L}$&log $\xi_{2}$&&(d.o.f)\\
       &  &  && & $\alpha_{4}$ &&&&\\
\hline
\multirow{4}{*}{4C\,74.26}   & \multirow{4}{*}{$0.61^{+0.09}_{-0.09}$} &  & \multirow{4}{*}{$1.70$} && $0.35^{+0.16}_{-0.16}$&  & & \multirow{4}{*}{$1.98^{+0.01}_{-0.01}$} &  \\
&& $< 0.01$\phantom{$^{00000000}$} && & $0.33^{+0.17}_{-0.16}$ & $< 0.1$\phantom{$0^{00000000}$} & $2.72$\phantom{$^{+0.00}$}  && 1.03 \\
       &  & -- & && $0.29^{+0.18}_{-0.17}$ & -- & $2.66^{+0.05}_{-0.10}$ & & (1272)\\
       &  &  & && $0.24^{+0.19}_{-0.19}$  & & & & \\
\hline
\multirow{4}{*}{Arakelian\,564}  & \multirow{4}{*}{$0.09^{+0.06}_{-0.09}$} &  & \multirow{4}{*}{$3.88$} && $0.69^{+0.14}_{-0.09}$ &  & & \multirow{4}{*}{$2.55^{+0.02}_{-0.01}$} &  \\
&& $0.44$\phantom{$^{+0.00}$} & &&  $0.62^{+0.17}_{-0.12}$ & $< 0.1$\phantom{$0^{0000000}0$} & $2.78$\phantom{$^{+0.00}$} && 1.05 \\
       &  & $3.46^{+1.27}_{-0.27}$ && & $0.56^{+0.20}_{-0.13}$ & -- &$1.65^{+0.25}_{-0.10}$ & & (\phantom{0}969)\\
       &  &  & && $0.48^{+0.24}_{-0.15}$ & & & &\\
\hline
\multirow{4}{*}{SWIFT\,J2127.4}  & \multirow{4}{*}{$1.50^{+0.11}_{-0.11}$} &  & \multirow{4}{*}{$2.54$} && $0.48^{+0.16}_{-0.14}$ &  &  & \multirow{4}{*}{$2.12^{+0.02}_{-0.01}$} &  \\
&& $< 0.01$\phantom{$^{00000000}$} & && $0.39^{+0.19}_{-0.16}$ & $< 0.1$\phantom{$0^{00000000}$} &$1.66$\phantom{$^{+0.00}$} && 1.00 \\
       &  & -- && & $0.32^{+0.21}_{-0.18}$ & -- & $1.65^{+0.15}_{-0.24}$ & & (\phantom{0}783)\\
       &  &  & && $0.19^{+0.26}_{-0.21}$ & & & &\\
\hline
\multirow{4}{*}{NGC\,7469}  & \multirow{4}{*}{$0.23^{+0.08}_{-0.09}$} & & \multirow{4}{*}{$0.91$} && $0.48^{+0.06}_{-0.07}$ &  &  & \multirow{4}{*}{$1.83^{+0.01}_{-0.01}$} & \\
&& $0.07$\phantom{$^{+0.00}$}  & && $0.39^{+0.06}_{-0.11}$ & $< 0.1$\phantom{$0^{00000000}$}  & $1.64$\phantom{$^{+0.00}$}  && 0.93 \\
       &  & $3.02^{+0.38}_{-0.23}$ & && $0.31^{+0.08}_{-0.14}$ & -- & $1.64^{+0.06}_{-0.13}$ & &(1076) \\
       &  &  & && $0.15^{+0.12}_{-0.21}$ & & & &\\
\hline
\multirow{4}{*}{NGC\,2992}  & \multirow{4}{*}{$1.71^{+0.06}_{-0.14}$} & & \multirow{4}{*}{$0.46$} && $0.42^{+0.10}_{-0.15}$ & &  & \multirow{4}{*}{$1.73^{+0.01}_{-0.01}$} &  \\
&&$< 0.01$\phantom{$^{00000000}$}  & && $0.36^{+0.08}_{-0.15}$ &  $< 0.1$\phantom{$0^{00000000}$} & $1.30$\phantom{$^{+0.00}$} && 1.03 \\
       &  &  -- && & $0.29^{+0.07}_{-0.16}$ & -- & $1.65^{+0.03}_{-0.26}$ & & (1660)\\
       &  &  && & $0.15^{+0.08}_{-0.21}$ & & & &\\
\hline
\multirow{4}{*}{MCG-5-23-16}  & \multirow{4}{*}{$2.40^{+0.02}_{-0.06}$} &  & \multirow{4}{*}{$4.11$} && $0.42^{+0.03}_{-0.02}$ & & & \multirow{4}{*}{$1.85^{+0.00}_{-0.00}$} & \\
&& $< 0.01$\phantom{$^{00000000}$} & && $0.31^{+0.03}_{-0.03}$ & $< 0.1$\phantom{$0^{00000000}$}  & $1.50$\phantom{$^{+0.00}$}  &&1.09 \\
       &  & -- && & $0.23^{+0.04}_{-0.03}$ & -- &$0.36^{+0.02}_{-0.04}$ & & (4692)\\
       &  &  && & $0.17^{+0.03}_{-0.03}$ & & & &\\
\hline
\multirow{4}{*}{IRAS\,18325}  & \multirow{4}{*}{$2.54^{+0.08}_{-0.08}$} &  & \multirow{4}{*}{$3.19$} && $0.57^{+0.04}_{-0.07}$ &  &  & \multirow{4}{*}{$2.40^{+0.01}_{-0.01}$} &  \\
&& $< 0.01$\phantom{$^{00000000}$} & && $0.43^{+0.06}_{-0.06}$ & $< 0.1$\phantom{$0^{00000000}$} & $1.57$\phantom{$^{+0.00}$} && 1.09 \\
       &  & -- && & $0.23^{+0.10}_{-0.07}$ & -- & $1.65^{+0.03}_{-0.05}$ & & (\phantom{0}980)\\
       &  &  && & $0.08^{+0.13}_{-0.08}$ & & & &\\
\hline
  \multicolumn{10}{@{}l@{}}{\hbox to 0pt{\parbox{180mm}{\footnotesize 
  \noindent 
  }\hss}}
\end{tabular}
}
\end{center}
\end{table*}

\end{document}